\documentclass[superscriptaddress,nofootinbib,preprintnumbers,twocolumn,10pt,aps,prd]{revtex4-2}
\usepackage{amsmath}
\usepackage{amssymb}
\usepackage{datetime}
\usepackage{graphicx}
\usepackage{hyperref}
\usepackage{cprotect}

\usepackage{blindtext}
\usepackage{hyperref}
\usepackage{amsmath,amssymb}
\usepackage{float}
\usepackage{microtype}
\usepackage{graphicx}
\usepackage{bm}
\usepackage{latexsym}
\usepackage{epsfig}
\usepackage{psfrag}
\usepackage{color}
\usepackage[dvipsnames]{xcolor}

\begin{document}


\title{New 511 keV line data provides strongest sub-GeV dark matter constraints}

 \author{Pedro De la Torre Luque}\email{pedro.delatorreluque@fysik.su.se}
 \affiliation{Departamento de F\'{i}sica Te\'{o}rica, M-15, Universidad Aut\'{o}noma de Madrid, E-28049 Madrid, Spain}
\affiliation{Instituto de F\'{i}sica Te\'{o}rica UAM-CSIC, Universidad Aut\'{o}noma de Madrid, C/ Nicol\'{a}s Cabrera, 13-15, 28049 Madrid, Spain}
 \affiliation{The Oskar Klein Centre, Department of Physics, Stockholm University, Stockholm 106 91, Sweden}

  \author{Shyam Balaji}
 \email{shyam.balaji@kcl.ac.uk}
 \affiliation{Physics Department, King’s College London, Strand, London, WC2R 2LS, United Kingdom}
\author{Joseph Silk}
\email{silk@iap.fr}
\affiliation{Institut d’Astrophysique de Paris, UMR 7095 CNRS \& Sorbonne Universit\'{e}, 98 bis boulevard Arago, F-75014 Paris, France}
\affiliation{Department of Physics and Astronomy, The Johns Hopkins University, 3400 N. Charles	Street, Baltimore, MD 21218, U.S.A.}
\affiliation{Beecroft Institute for Particle Astrophysics and Cosmology, University of Oxford, Keble	Road, Oxford OX1 3RH, U.K.}

\smallskip
\begin{abstract}
We explore the $511$~keV emission associated to sub-GeV dark matter (DM) particles that can produce electron-positron pairs and form positronium after thermalizing. We use $\sim16$~yr of SPI data from INTEGRAL to constrain DM properties, including the full positron propagation and losses, and the free electron density suppression away from the Galactic plane. We show that the predicted longitude and latitude profiles vary significantly for different DM masses, unlike previous assumptions, and obtain the strongest limits on sub-GeV DM (from the MeV to a few GeV) so far, excluding cross-sections down to $\langle \sigma v \rangle \lesssim10^{-32}$ cm$^3$ s$^{-1}$ for $m_{\chi}\sim1\,\text{MeV}$ and $\langle \sigma v \rangle \lesssim10^{-26}$ cm$^3$ s$^{-1}  $ for $m_{\chi}\sim5\,\text{GeV}$ and lifetimes up to $\tau \gtrsim 10^{29}\, \textrm{s}$ for $ m_{\chi}\sim1\,\text{MeV} $ and $\tau \gtrsim 10^{27}\,\textrm{s}$ for $m_{\chi}\sim5$~GeV for the typical Navarro-Frenk-White DM profile. Our derived limits are robust within a factor of a few due to systematic uncertainties.

\end{abstract}
\maketitle

\section{Introduction}
The 511 keV $\gamma$-ray line from the galactic bulge indicates $e^+ e^-$ annihilation into $\gamma\gamma$ via positronium in the interstellar medium (ISM). This line has been measured by SPI on INTEGRAL \citep{Siegert:2015knp} and COSI \citep{Kierans:2019aqz}, and reviewed in~\cite{Prantzos:2010wi, kierans2019positron, Siegert_2023}. The line has disk and bulge components, with the bulge component being narrow, bright and centered at the galactic center (GC), with a flux of $\sim 10^{-3}$ cm$^{-2}$ sec$^{-1}$ \citep{Siegert:2015knp}. This has been used to constrain possible positron sources, such as isotopes from stars \citep{Prantzos:2010wi, Bartels:2018eyb}, low-mass $X$-ray binaries \citep{Bartels:2018eyb} or neutron star mergers \citep{Fuller:2018ttb}. More recently stellar templates for the nuclear stellar bulge and a boxy bulge were shown to have good matches with the observed signal~\cite{Siegert:2021trw}.

The bulge positrons could also come from dark matter (DM), which is abundant near the GC. This possibility has been explored, at least, since \cite{Boehm:2003bt}. However, not all DM models work. For example, DM decay does not fit the signal shape \cite{Vincent:2012an,Ascasibar:2005rw}. Moreover, the positron injection energy must be around MeV scale \cite{Beacom:2004pe, boehm2006revisiting, Beacom:2005qv, Boehm_2009}, which implies that the DM mass must be below some tens of MeV, unless there are intermediate steps in the annihilation. Such a low DM mass was thought to be inconsistent with cosmology, but recent studies have shown that it can be compatible with observations if there is also neutrino injection from DM annihilation in the early universe \cite{Escudero:2018mvt,Sabti:2019mhn}. We use these new findings to motivate study of light DM scenarios that may produce the 511 keV line.

\begin{figure}[t!]
\centering
\includegraphics[width=0.99\linewidth]{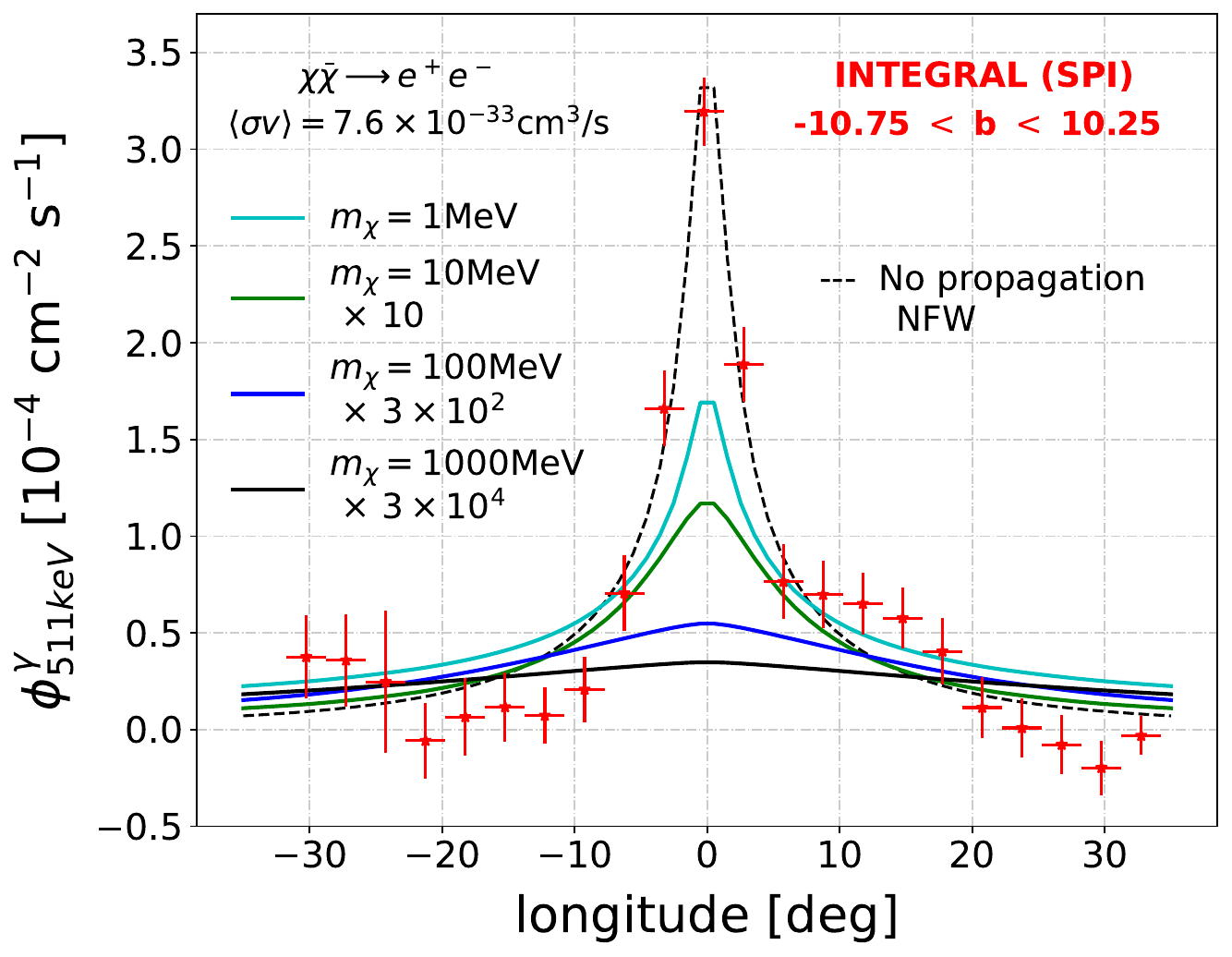}
\caption{Diffuse $511$~keV emission longitude profile in the latitude bin $-10.75^\circ<b<10.25^\circ$ from direct DM annihilation into electron-positron pairs. We show DM masses ($m_\chi$) ranging from $1$ (cyan), $10$ (green), $100$ (blue) and 1000~MeV (black) respectively, and compare to the expected signal with no propagation (dashed). To facilitate the comparison with SPI data (shown in red), the emission for each mass is scaled by the factor indicated in the legend. We consider a typical NFW DM distribution.} 
\label{fig:Ann_MassComp_ee}
\end{figure}

\cite{Vincent:2012an} showed that the DM annihilation rate needed to explain the 511 keV line emission in the bulge (from $\sim 8$ years of SPI data in a 5 keV bin), was approximately
\begin{align}
\langle\sigma v\rangle_{511}
\simeq 5\times 10^{-31} \Big(\frac{m_\chi}{3~\text{MeV}}\Big)^{2}\, \frac{\text{cm}^3}{\text{sec}},
\label{eq:fit511}
\end{align}
for a Navarro-Frenk-White (NFW) profile DM density distribution for the best fit parameters in~\cite{Vincent:2012an}. 

However, the previous studies have assumed that positrons follow the DM distribution and have ignored positron propagation in the Galaxy, that depends on their energy, $E$ (with $D(E) \propto E^{\delta}$, where $D$ is the diffusion coefficient and $\delta>0$ means that higher energy positrons diffuse faster). This assumption was adopted since energy losses at MeV energies are dominant \cite{Strong_1998}, but diffusion is still important, and leads to a $511$~keV profile that changes with the DM mass, as we show in Fig.~\ref{fig:Ann_MassComp_ee}, where we also compare with the profile expected from a NFW DM density distribution without any propagation. Moreover, the latest data from AMS-02 \cite{AMS_gen} on secondary cosmic-rays (CRs) suggest a faster diffusion below the GeV scale \citep{Weinrich_2020, Luque:2021nxb, Luque:2021ddh, delaTorreLuque:2022vhm}. Also, the interaction of charged particles with turbulence in the interstellar medium (ISM) causes energy exchange and diffusive reacceleration \citep{1995ApJ...441..209H, Strong_1998, 2019Galax...7...49C}, which boosts low-energy particles to higher energies, making them diffuse faster. Therefore, we evaluate the 511 keV profile from the distribution of propagated positrons injected by DM, instead of assuming that it follows the DM distribution. In fact, some recent works \citep{DelaTorreLuque:2023huu, DelaTorreLuque:2023nhh, DelaTorreLuque:2023olp}, have shown the importance of adding propagation effects in constraining sub-GeV DM.

Here, we perform a detailed evaluation of the $511$~keV line emission from DM and utilize the non-compatibility of these signals with the observed $511$~keV line morphology, to constrain the DM annihilation cross-section and decay lifetime for the kinematically accessible channels. In doing so, we highlight three major novelties. i) We include the transport of positrons when calculating the $511$~keV emission in the Galaxy. We show that the diffusion of positrons changes their spatial morphology significantly. ii) We include the fact that the free electron density drops quickly away from the Galactic Plane, which suppresses the latitude profile flux of the $511$~keV emission. This is necessary for any positron source that could produce $511$~keV radiation. iii) We derive strong limits on the lifetime of decaying DM. Remarkably, we find that the limits from the full SPI dataset are the most constraining ones below DM masses of $\sim1$~GeV, a mass range of interest given low sensitivity in DM direct detection experiments to light DM \citep{Cirelli:2023tnx, DelaTorreLuque:2023olp}.

\section{Evaluation of the 511~keV signal} 
\label{method}
The literature assumes that the distribution of diffuse positrons from DM decay/annihilation follows the parent DM Galactic distribution. However, positrons can propagate and diffuse far from their source, changing their distribution and the profile of the observed diffuse $511$~keV line in the Galaxy~\cite{JeanP_2009}. Therefore, we evaluate the DM-induced $511$~keV line flux considering the full propagation of positrons, including all sources of energy losses and other effects, such as reacceleration~\cite{seo1994stochastic, osborne1987cosmic}, triplet pair production~\cite{Gaggero:2013eik} and in-flight annihilation~\cite{Heitler:1936jqw}. We calculate the steady-state solution for the distribution of positrons injected by DM (similar to~\cite{Calore:2021lih}) and integrate it over all possible energies.

We compare our results with more than $16$~years of data from INTEGRAL/SPI, which offered the latitudinal and longitudinal profiles of the diffuse 511 keV line emission with unprecedented quality~\cite{Siegert_2019}.
\newline
\subsection{Positron propagation}
We compute the distribution and energy spectra of the positrons produced by DM annihilation/decay in the Galaxy with a customized version~\cite{de_la_torre_luque_2023_10076728} of the {\tt DRAGON2} code~\cite{DRAGON2-1, DRAGON2-2}, a CR propagation code that numerically solves the full diffusion-advection-loss equation for charged particles in the Galactic environment~\cite{Ginz&Syr}. We use the same expression for the diffusion coefficient and propagation parameters as in~\cite{DelaTorreLuque:2023olp}, which, at energies below $300$~GeV, takes the form
\begin{equation}
    D(E) = D_0 \beta^{\eta}\left(\frac{E}{4 GeV}\right)^{\delta} \,\, ,
\end{equation}
where $\delta=0.49$, $\eta=-0.75$, $D_0=1.02\times10^{29}$~cm$^2$/s and $\beta$ is the ratio of the particle speed to the speed of light. This setup allows us to reproduce the current experimental CR data with very good accuracy, as broadly discussed and shown in Appendix A of~\cite{DelaTorreLuque:2023olp}. These simulations include all the relevant sources of energy loss (namely, synchrotron, inverse Compton, bremsshtralung ionization and Coulomb losses), inelastic interactions, triplet pair production~\cite{Gaggero:2013eik} and in-flight annihilation of positrons. We show in Appendix~\ref{sec:AppTimeScales} a comparison of the relevant timescales of the simulation.

We simulate the channels leading to the production of $e^{\pm}$ final states through the processes: $\chi\overline{\chi}\rightarrow \mu^+\mu^-,\pi^+\pi^-,e^+e^- $ for DM annihilation and $\chi\rightarrow \mu^+\mu^-,\pi^+\pi^-,e^+e^- $ for DM decay.
The injection spectra $dN_e/dE_e$ for each channel are calculated following~\cite{Cirelli:2020bpc}.
We compute the signals from light DM from $m_{\chi} = m_{i}$ for annihilation and $m_{\chi} = 2 m_{i}$ for decays, up to $m_{\chi}=5$~GeV (where $i=e,\mu,\pi$), taking $3$ masses per decade and a NFW DM distribution~\cite{Navarro:1995iw}, as a baseline.

Simulations are performed with a spatial resolution of $\sim150$~pc, which is sufficiently fine for comparisons with SPI data and for the smooth NFW profile. We simulate electron-positron signals from DM annihilation and decay in the range of kinetic energies from $100$~eV to $10$~GeV, with an energy resolution of 5\%. We need to consider such low energies for reasons explained below. Lowering the minimum energy or the spatial resolution does not impact our results significantly. We ensure convergence of the simulations using a variable time step (see Sect.~4.5 of~\cite{Evoli2017jcap}), with a minimum time step of $0.1$~kyr and a maximum time step of $64$~Gyr, that cover the relevant timescales involved.

\subsection{511 keV emission}
Given the propagated and integrated flux of diffuse positrons in the Galaxy as a function of 3D position $(x,y,z)$, $\phi_{\rm e}(x,y,z)$, we calculate the emission of $511$~keV $X$-rays from a given direction by integrating over the line of sight, as
\begin{equation}
\frac{d\phi_{\gamma}^{511}}{d\Omega}=2k_{ps}\int ds\,s^{2}\frac{\phi_e(x_{s,b,l},y_{s,b,l},z_{s,b,l})}{4\pi s^{2}},
\end{equation}
where $k_{ps}=1/4$ is the fraction of positronium decays corresponding to (singlet) para-positronium states) contributing to the  $511$~keV line signal, $\phi_{\rm e} = \int \frac{d\phi_{\rm e}}{dE} dE$ is the energy-integrated flux of positrons, $d\Omega=dl db \cos b$ is the solid angle element with $l$, $b$ and $s$ denoting the Galactic longitude, latitude and distance $s$ (along the line of sight) from the Earth.

We note that the probability of producing positronium away from the Galactic Plane depends on the positron density injected from DM and the ambient free electron density. Hence, we apply a scaling relation to the $511$~keV profiles, following the vertical distribution of free electron density in the Galaxy. This is motivated by the fact that SPI observations of the latitude profile show a similar variation of the $511$~keV emission as the one of the free electron density (see Appendix~\ref{sec:Scaling} and Fig.~\ref{fig:Lat_Scal}). This scaling leads to a more conservative but realistic evaluation of the predicted $511$~keV emission.
For the distribution of free electrons, we use the NE2001 model~\cite{cordes2003ne2001, cordes2003ne2001i}, which contains the modelling of several Galactic sub-structures and is derived from a combined fit of pulsar dispersion measurements, temporal and angular broadening of radio pulses, and intensity measurements. We consider only the vertical dependence of the electron density, since the free electron density is always much higher than the diffuse positron density in the Galactic disk. This has usually been assumed true in all of the Galaxy (see, e.g. \cite{KeithHooper_511Lat, Calore:2021lih}), however a different rate of positronium production is expected out of the Galactic plane, because of the steep reduction in free electron density when moving away from the Galactic plane. This scaling has an appreciable effect in the predicted latitude profiles, but has almost no effect on the predicted longitude profiles of the $511$~keV line emission.

We adopt the scaling predicted by the NE2001 model, applying the correction proposed in~\cite{Gaensler_2008}, where the authors found that the height of the thick disk roughly doubles to $\simeq 2$~kpc:
i) a thick disk with large Galactocentric scale height (whose scaling follows a $\exp[-|z|/H_1]$ relation, where $z$ is the Galactocentric height and the scale $H_1$ was found to be $1$~kpc), ii) a thin, annular disk in the inner Galaxy (whose scaling decays as $\exp[-|z|/H_2]$, with $H_2\simeq 140$~pc), and iii) a GC component.
We take this scaling to be maximal at the center of the Galaxy and exponentially decays like $\exp[-|z|/H_1]+\exp[-|z|/H_2]$ at a height $z$, as shown in Fig.~\ref{fig:Lat_Scal}.

\section{Sub-GeV dark matter constraints from SPI data} 
\label{sec:results} 
In Fig.~\ref{fig:Ann_MassComp_ee}, we depict how predicted $511$~keV line longitude profiles vary with different DM masses, specifically when directly annihilating into $e^+e^-$ final states. Results for DM masses of $1$, $10$, $100$, and $1000$~MeV are compared to SPI data. A decreasing DM mass results in an increase of the signal normalization and a peakier profile around the GC. This effect is due to lower-energy injected positrons covering shorter distances before thermalization or annihilation, influenced by higher gas density toward the GC. 
Lower DM masses match better with SPI's longitude profile shape. However, the NFW profile fails to replicate the peakiness indicated by SPI measurements for DM masses above the electron mass. Only a highly peaked distribution like the Moore profile \cite{Moore_1999} ($\gamma=1.5$) provides a satisfactory fit, as shown in Fig.~\ref{fig:Ann_DMProfs_ee}. This emphasizes the importance of data away from central longitudes for deriving stringent DM bounds.

\begin{figure*}[t!]
\includegraphics[width=0.49\linewidth]{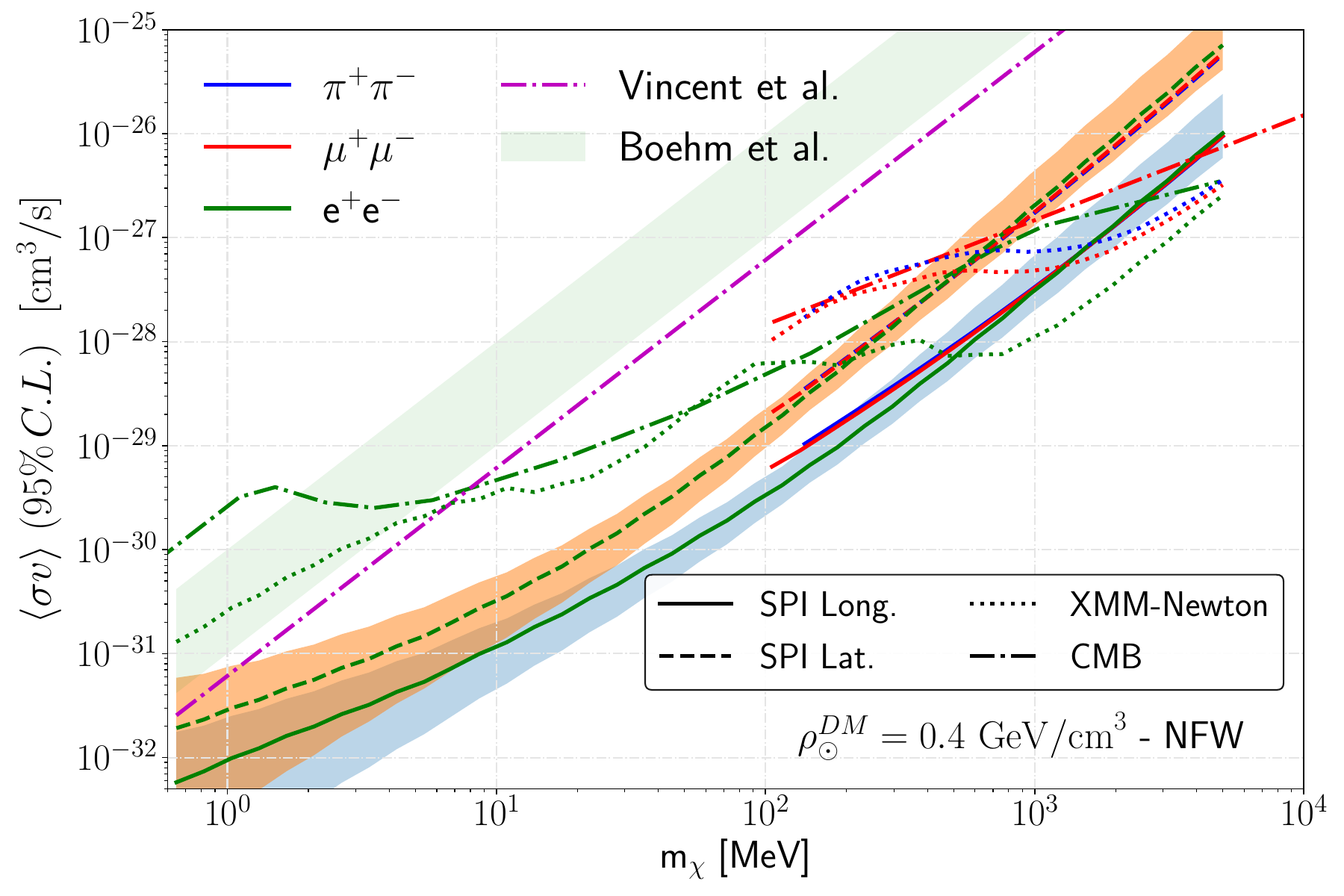}
\includegraphics[width=0.49\linewidth]{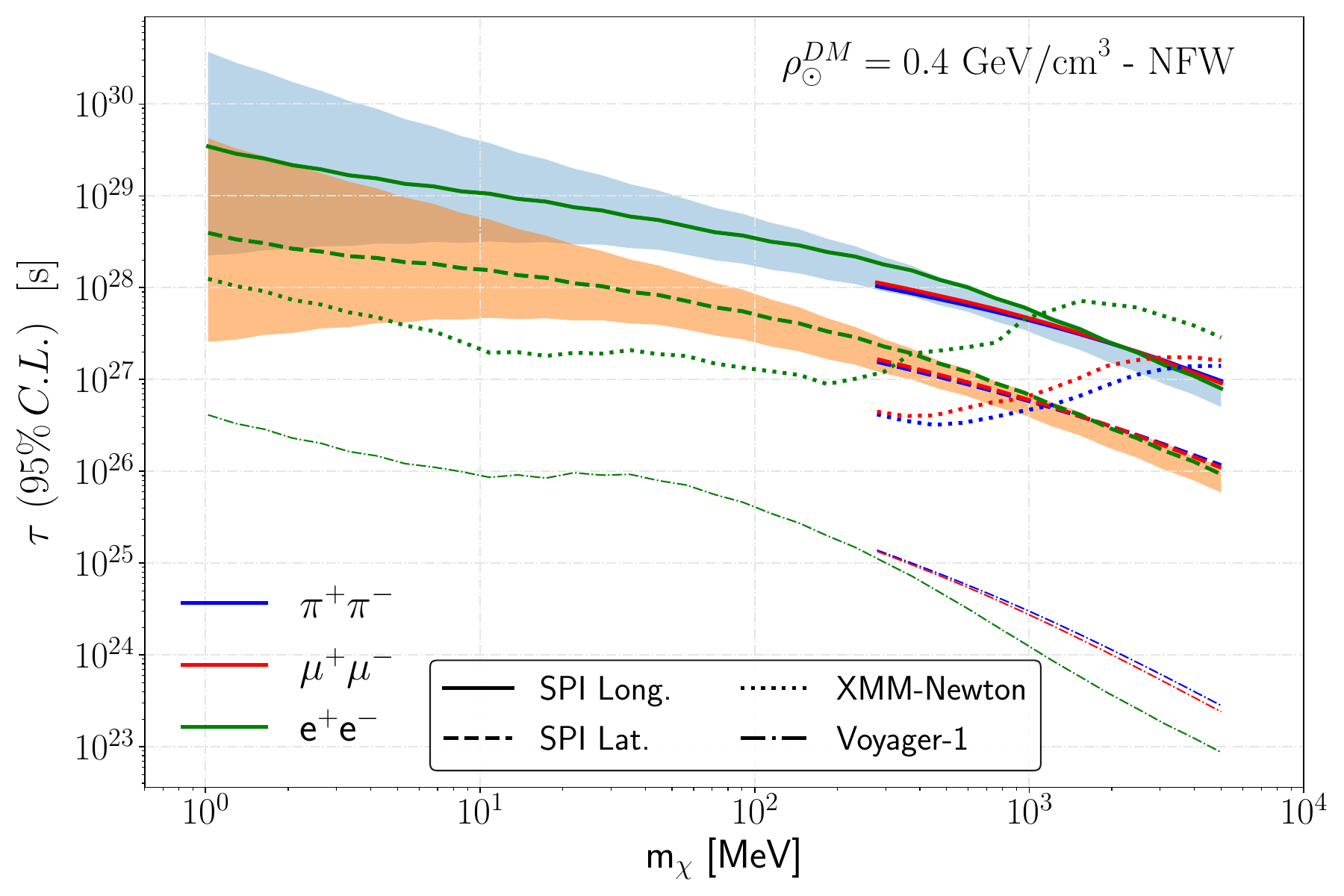}
\caption{Comparison of the $95\%$ confidence bounds on annihilating (left panel) or decaying (right panel) DM derived in this work (thick dashed for {\sc SPI Lat} and solid lines for {\sc SPI Long}, respectively) with other existing constraints. We show the CMB bounds from Slayter~\cite{Slatyer:2015jla} and Lopez-Honorez et al.~\cite{Lopez-Honorez:2013cua} (dot-dashed line), the bounds from {\sc XMM Newton}~\cite{DelaTorreLuque:2023olp} (dotted line), the previous 511 keV line bounds from Vincent et. al~\cite{Vincent:2012an} (red dashed line) and from {\sc Voyager 1}~\cite{DelaTorreLuque:2023olp} (dot-dashed line). In both panels, we show the $\pi^+\pi^-$ (blue), $\mu^+\mu^-$ (red) and $e^+e^-$ (green) channels, respectively. Uncertainties from the propagation setup employed are shown as an orange and a blue bands, for the latitude and longitude profiles, respectively.}
\label{fig:Limits}
\end{figure*}

Similar plots for other annihilation channels ($\mu^+\mu^-$, $\pi^+\pi^-$, and direct $e^+e^-$) are presented in Fig.~\ref{fig:Ann_Prof_pimu}, including the predicted longitude and latitude profiles for various DM masses. For latitude profiles, scaling the free electron density yields a profile similar to SPI's, even at high ($\sim$ GeV) DM masses. 
Considering positron annihilation in flight is crucial, impacting the distribution above or below the Galactic plane, particularly reducing emission in the plane with lower DM masses.

Additionally, as confirmed by other studies~\cite{Boehm:2003hm, Vincent:2012an}, decaying DM produces a flat profile inconsistent with observations, evident in Fig.~\ref{fig:Dec_MassComp_ee}. 

We show the bounds from SPI data and compare them to other DM constraints in Fig.~\ref{fig:Limits}, for DM annihilation (left panel) and decay (right panel). To derive these limits, we use a simple $\chi^2$ fit to the SPI data, using the {\sc curve-fit} {\sc Python} package, and take the $2\sigma$ uncertainty bands. We show the constraints from the SPI latitude profile as a dashed curve (``SPI Lat") and the SPI longitude profile as a solid curve (``SPI Long"). We use colour to distinguish the channel of annihilation or decay: $e^+e^-$ (green), $\mu^+\mu^-$ (red) and $\pi^+\pi^-$ (blue). We restrain ourselves to computing the limits independently, since we expect that combining both datasets will not offer noticeable improvement and will be dominated by longitude data.
In the left panel, we also compare with the limits from the $e^\pm$ injection from DM annihilation on the CMB anisotropies, from Slatyer~\cite{Slatyer:2015jla} and Lopez-Honorez et al.~\cite{Lopez-Honorez:2013cua}, shown as dot-dashed lines (``CMB"). These limits are based on the fact that the injection of charged particles by DM during the cosmic Dark Ages must increase the residual ionization fraction, thus altering the anisotropies of the CMB. The CMB limits are weak for decaying light DM \cite{Liu:2023fgu, Liu:2023nct}, unlike the case of annihilating DM. This is because DM clusters at redshifts $z \lesssim 100$, which increases the DM annihilation rate and the $e^\pm$ injection, while the decay rate stays the same. Since the CMB bounds are weak for decaying DM, we show the stronger bound from~\cite{DelaTorreLuque:2023olp} (labelled ``Voyager-1") as a dot-dashed curve in the right panel. This limit comes from the fact that decaying DM results in an $e^\pm$ cascade that can be observed by {\sc Voyager 1}, without the solar modulation effect of the heliosphere, due to the satellite's distance from the solar system.

We show the strong bounds from~\cite{DelaTorreLuque:2023olp}, labelled as ``XMM-Newton" in both panels, as dotted curves. These limits are derived from the fact that $e^\pm$ products from an exotic DM injection of $e^{\pm}$ may generate bremsstrahlung radiation and upscatter the low-energy Galactic photon fields via the inverse Compton process, producing a broad emission from $X$-ray to $\gamma$-ray energies observable by {\sc Xmm-Newton}, which measured $X$-rays in galactocentric rings around the GC. These limits include best-fit CR propagation and diffusion parameters and represented the strongest astrophysical constraints for this mass range of DM of $1$~MeV to a few GeV and surpass cosmological bounds across a wide range of masses.

We show the bounds from~\cite{Vincent:2012an}, also using $511$~keV data from SPI after a shorter $\sim8$~yr of operation. Our limits are stronger because their constraint used the spectrum of the emission at 511 keV, instead of the full profile. This, combined with the fact that we use the most constraining observations (from high longitude), explains the relative strength of our constraints.

In fact, we see that the SPI longitude profile provides the strongest limit on annihilating DM, excluding cross-sections of $\langle \sigma v\rangle\simeq10^{-32}\,\textrm{cm}^3\textrm{s}^{-1}$ at masses of around an MeV, increasing to around the thermal relic values of $\langle \sigma v\rangle\simeq10^{-26}\,\textrm{cm}^3\textrm{s}^{-1}$ at masses of over several GeV. This SPI longitude profile only gets surpassed by {\sc Xmm-Newton} $X$-ray bounds at masses $\gtrsim 700$ MeV. The CMB bounds only become more stringent than the SPI latitude profile limits at masses in excess of a few tens of GeV. 
Strikingly, limits derived from the SPI latitude profile are only marginally weaker (less than an order of magnitude across the entire DM mass range) than the SPI longitude profile limits. This was also found in~\cite{Calore:2021lih}. Furthermore, this bound also surpasses the other DM limits across most of the mass range considered, except {\sc XMM-Newton} at masses over a few hundreds of MeV and CMB bounds at slightly higher mass. Finally, we also compare our limits with those derived from~\cite{Boehm_2004} from final state radiation emission.

From the right panel of Fig.~\ref{fig:Limits}, we see once again that the SPI longitude profile provides the strongest limit on decaying DM, excluding lifetimes of $\tau \simeq 10^{29}\,\textrm{s}$ at masses of around an MeV decreasing to around $10^{27}\,\textrm{s}$ at masses of over a GeV. This SPI longitude profile gets surpassed by {\sc Xmm-Newton} $X$-ray bounds at masses $\gtrsim 1$ GeV. Limits derived from the SPI latitude profile are about an order of magnitude weaker than the SPI longitude profile limits across the entire mass range. The next strongest limit, coming from {\sc Voyager 1} is significantly weaker (about three orders weaker than the SPI longitude profile limits). This illustrates the power of the $511$~keV line in constraining sub-GeV DM. We remark that including the free electron density scaling changes our latitude bounds by a factor of $2$-$3$, leading to a more conservative estimate.

\section{Discussion and conclusion} 
\label{sec:Conclusion}
This study investigates sub-GeV dark matter (DM) particles annihilating or decaying into standard model (SM) particles, leading to an electron-positron cascade and positronium bound-state formation. About one-fourth of these states decay into 511 keV photons, detectable by the SPI spectrometer on the INTEGRAL satellite. SPI longitude and latitude profiles are used to establish new limits on DM properties, including mass, annihilation cross-section, and decay lifetime. We enhance previous analyses by considering two previously overlooked effects: positron diffusion and propagation in the interstellar medium, and the decrease in free electron density away from the Galactic plane. These considerations significantly impact the 511 keV photon line's shape and intensity, influencing derived DM constraints. Additionally, this study presents the first constraints on decaying DM using the SPI dataset. Assuming a NFW density distribution, our limits are the strongest across the sub-GeV DM mass range, ruling out thermally-averaged cross-sections in the range $10^{-32}$ cm$^3$s$^{-1} \lesssim \langle \sigma v\rangle \lesssim 10^{-26}$ cm$^3$s$^{-1}$ and decay lifetimes of $10^{29}\,\textrm{s}\lesssim \tau \lesssim 10^{27}\,\textrm{s}$, surpassing many cosmological and astrophysical limits.
However, uncertainties persist: On the one hand, the total systematic uncertainties of the SPI data points are not totally under control, since they depend on some template models for the shape of the signal. The fact that these constraints are so sensitive to the high longitudes can be particularly affected by these systematic uncertainties, as pointed out in~\cite{DelaTorreLuque:2023huu}, given that there are data points indicating no flux detected at these longitudes. In that paper, we found that constraining ourselves to the inner $20^{\circ}$ instead (expected to be less affected by the detector's systematic uncertainties) leads to a weaker limit by a factor of $2$.
On the other hand, we expect that uncertainties in the CR propagation setup affect the predicted flux of positrons from DM. \cite{DelaTorreLuque:2023olp} demonstrated that the main factors affecting the flux of positrons produced from sub-GeV DM are the halo height and the Àlfven speed (dictating the level of reacceleration). Therefore, we have built a pessimistic and optimistic propagation setups to compute the uncertainty bands shown in Fig.~\ref{fig:Limits}. The pessimistic setup features the scenario with very slow diffusion, where the halo height is set to $H=3$~kpc (given that the lower the halo height, the lower the positron flux~\citep{DelaTorreLuque:2023olp}, we take a very small value compared to the typical values obtained in CR analyses), keeping the ratio of the normalization of the diffusion coefficient to the halo height ($H$/$D_0$) to the value that reproduces CR secondary ratios at Earth. In addition, we assume no reacceleration (i.e. $V_A = 0$~km/s). In the optimistic case, we set  $V_A = 40$~km/s and $H=16$~kpc. As we observe from Fig.~\ref{fig:Limits}, at low masses, uncertainties in reacceleration make our limits vary by up to an order of magnitude, while at high masses the effect of reacceleration is minor and uncertainties in the halo height make our limits uncertain by a factor of up to $2$-$3$, compatible with what was also found in~\citep{DelaTorreLuque:2023olp}.
Effects like strong convection or winds near the Galactic Center could also alter the positron spectra~\citep{Bartels:2017dpb}, but accurate estimation of this with current data is challenging. 
On top of this, we have checked that these limits do not change by more than a factor of a few using a cuspier DM distribution than the NFW. An Einasto profile is expected to slightly strengthen these limits, given that it results in more DM density at intermediate longitudes. Following the same reasoning, the use of a more cored profile, like the Isothermal one, could worsen the limits. However, even with these uncertainties included, our results show that the $511$~keV line still provides the strongest astrophysical bounds so far, especially for DM decay across the full MeV range.

Future data from the COSI~\citep{Siegert_2020} with improved spatial resolution would provide insights into processes and structures producing the 511 keV line in the Milky Way.
Additionally, testing these DM scenarios by characterizing diffuse MeV-scale $\gamma$-ray emission from the Milky Way halo, using proposed $\gamma$-ray telescopes like AMEGO \citep{mcenery2019allsky} or e-ASTROGAM \citep{De_Angelis_2018}, could provide useful background models and further strengthen bounds on sub-GeV DM.

\textbf{Acknowledgements} \newline
We are grateful to Pierluca Carenza, Marco Cirelli, Daniele Gaggero, Dan Hooper, Jordan Koechler, Tim Linden, Pierrick Martin and Thomas Siegert for useful discussions. SB is supported by the STFC under grant ST/X000753/1. PDL is supported by the European Research Council under grant 742104 and the Swedish National Space Agency under contract 117/19. This project used computing resources from the Swedish National Infrastructure for Computing (SNIC) under project Nos. 2021/3-42, 2021/6-326, 2021-1-24 and 2022/3-27 partially funded by the Swedish Research Council through grant no. 2018-05973.


\appendix

\section{Height scale of the free electron density}
\label{sec:Scaling}
\begin{figure*}[b!]
\begin{center}
\includegraphics[width=0.47\linewidth, height=0.255\textheight]{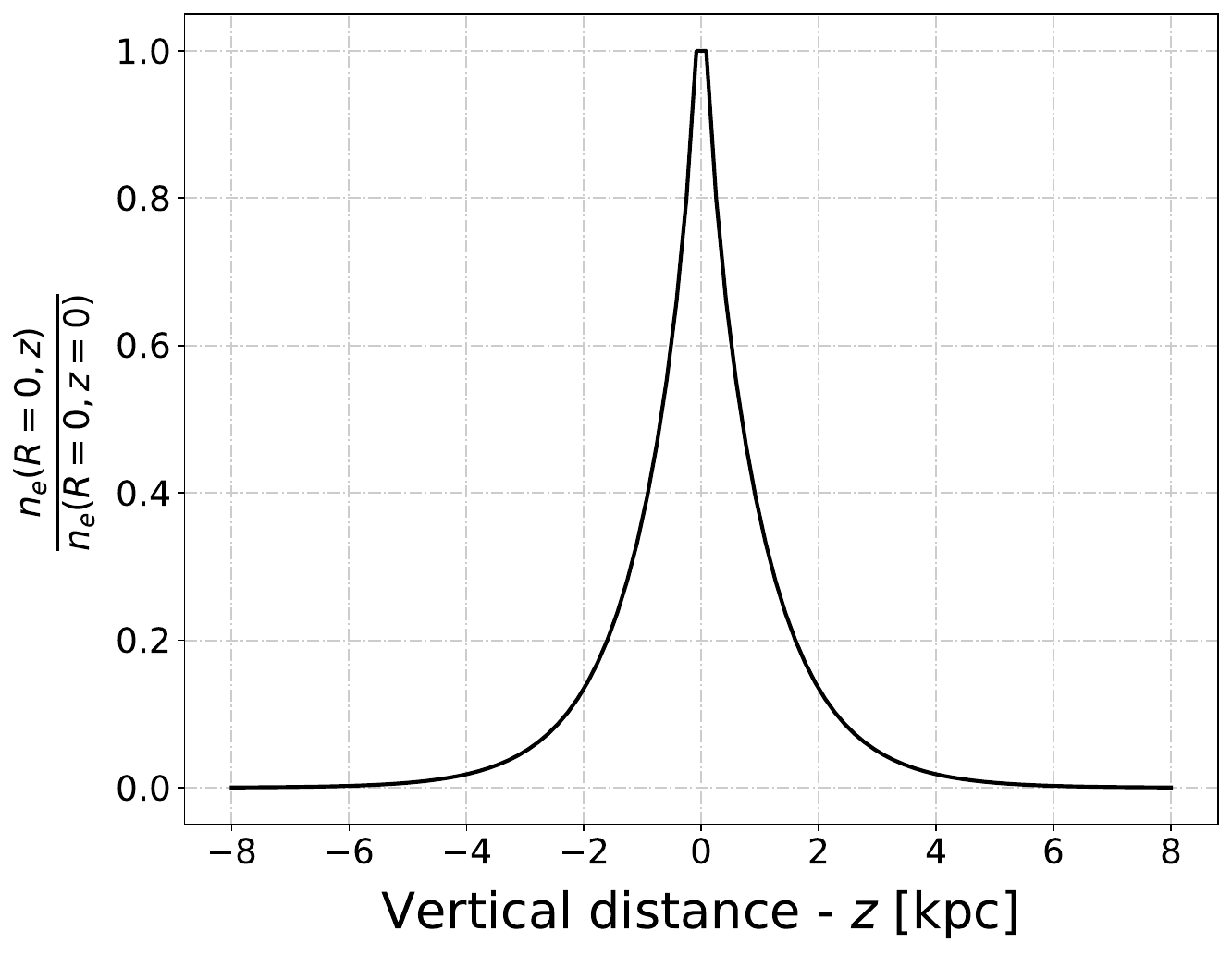}
\caption{Vertical profile of the free electron density normalized at the Galactic Center as a function of height above the Galactic Plane $z$. We take this distribution from the NE2001 model~\cite{cordes2003ne2001, cordes2003ne2001i}, as indicated in the text.}
\label{fig:Lat_Scal}
\end{center}
\end{figure*}

As explained above, in order to consider that different regions of the Galaxy will have a different rate (probability) of positronium formation, that also depends on the distribution of the ambient (free) electrons, we consider a scaling of the $511$~keV line emission following the NE2001 model~\cite{cordes2003ne2001, cordes2003ne2001i} for the free electron density distribution.
In particular, we consider that the disk of the Galaxy will always contain a much larger density of free electrons in the ISM gas than diffuse positrons (i.e. there will be always more electrons available than positrons to form positronium states, thus being ``saturated'', and the probability of positronium formation will mainly depend on the diffuse positron distribution). However, this assumption must be broken well above and below the disk, since the density of free electrons falls very quickly as the galactic height, $z$, is increased, as shown by Fig.~\ref{fig:Lat_Scal}. In particular, in this figure we depict the electron density distribution, at the center of the Galaxy, as a function of the Galactic height, normalized to the electron density at the GC position. This is the scale factor that we apply to our evaluations of the $511$~keV line. This scaling is particularly important for very low DM masses, when the positrons lose energy very fast in the disk (but not outside the disk, since the density of gas producing these losses is much lower) and the fraction of positrons annihilated in flight (directly proportional to the gas density where the positrons propagate) becomes relevant.

This is also motivated by the fact that, as we observe from the SPI measurements of the latitude profile, the intensity of the $511$~keV line decays by around one order of magnitude at $10^{\circ}$ above or below the disk. Assuming the the height of the halo is around $4$-$10$~kpc, this means that with every $\sim0.36$-$0.9$~kpc, the intensity of the line decays by one order of magnitude, similar to the factor $\gtrsim6$ that the free electron density decays every $\sim$kpc above or below the disk.
While the predicted latitude profiles from DM appear to be quite flat (compared to the SPI data) for all the masses explored here, using this scaling we find that the predicted latitude profiles become much more similar to the shape of the SPI measurements. However, we remark that the longitude profiles predicted are roughly unaffected by this scaling.

\section{Relevant timescales}
\label{sec:AppTimeScales}

\begin{figure*}[h!]
\centering 
\includegraphics[width=0.5\linewidth]{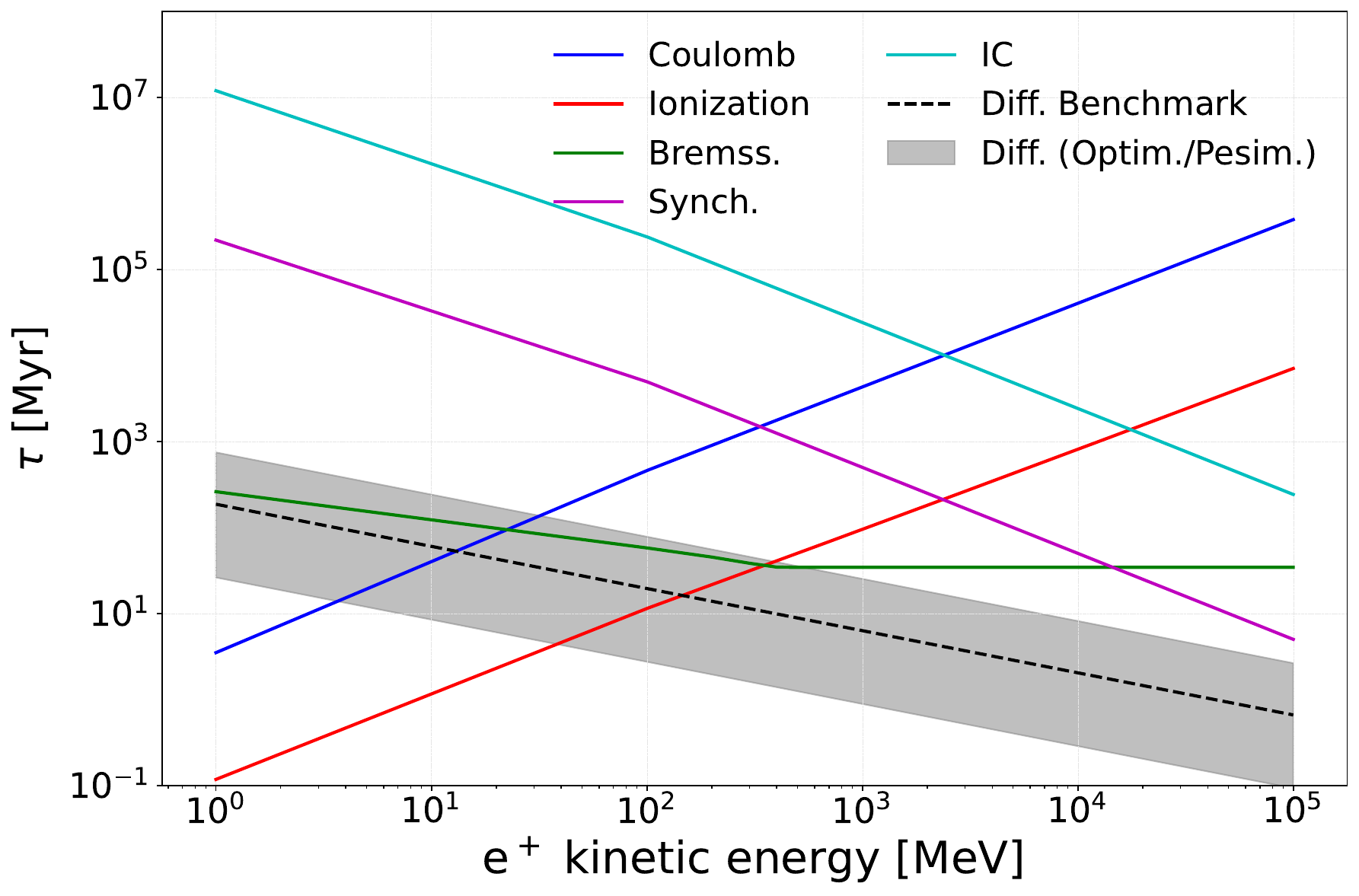}
\caption{Main timescales involved in the propagation of electrons and positrons generated by sub-GeV DM particles, for a gas density of $1$~cm$^{-3}$ composed by H and a $10\%$ of He. Gas
densities. The magnetic field strength here is set to $5$~$\mu$G and the energy density of the interstellar radiation fields is taken from~\cite{Evoli_2020}. The grey band indicates the difference in diffusion time in the Optimistic and Pessimistic setups, used to calculate uncertainties in the DM bounds.}
\label{fig:TimeScales}
\end{figure*}

Our calculations include all the relevant Galactic CR propagation processes in a broad energy range. For signals from sub-GeV DM, it is crucial to account for diffusion, reacceleration, and energy losses, that become dominant below a few tens of MeV. The most relevant cooling processes at these energies are Coulomb and ionization losses, although we include in our simulation synchrotron, bremsstrahlung, inverse Compton and catastrophic losses
(annihilation) as well.
A comparison of the relevant timescales is illustrated in Fig.~\ref{fig:TimeScales}, where we show a band around the benchmark diffusion time (i.e. diffusion time of particles with our benchmark diffusion setup) to indicate the difference between the optimistic and pessimistic scenarios. Moreover, cooling is much faster than diffusion. We note that the average distance travel by a positron injected at $10$~MeV (which can be calculated as $\langle l_{\text{travel}} \rangle \approx \sqrt{2D\tau}$, being $\tau$ the dominant timescale) is larger than $100$~pc in a $1$~cm$^{-3}$ gas, without accounting for wind advection or reacceleration (that can boost $\langle l_{\text{travel}} \rangle$ to even more than $500$~pc).

\section{511 keV emission from common Galactic dark matter density distributions}
\label{sec:Profiles_DM_Dist}
\begin{figure*}[h!]
\includegraphics[width=0.495\linewidth]{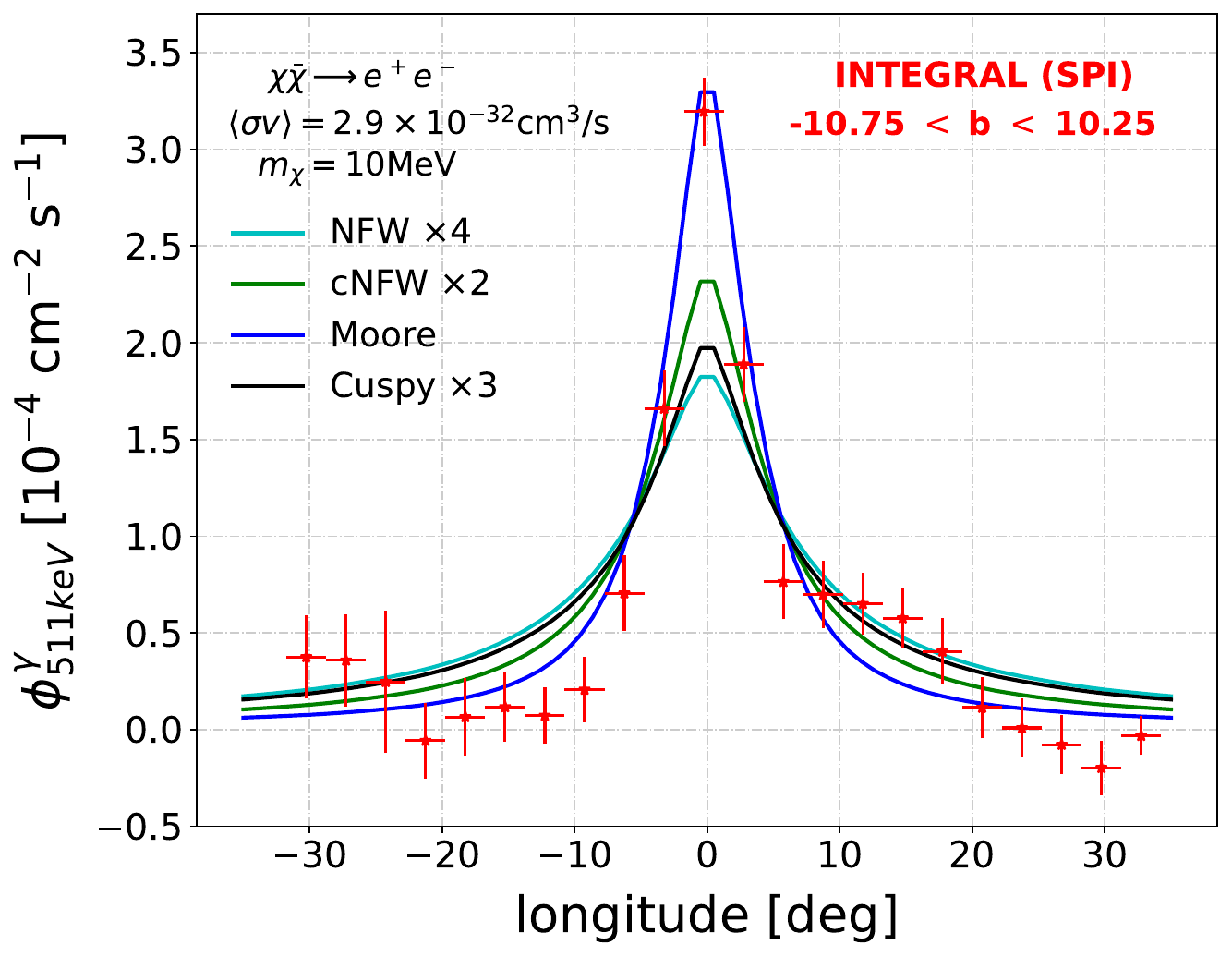}
\includegraphics[width=0.495\linewidth]{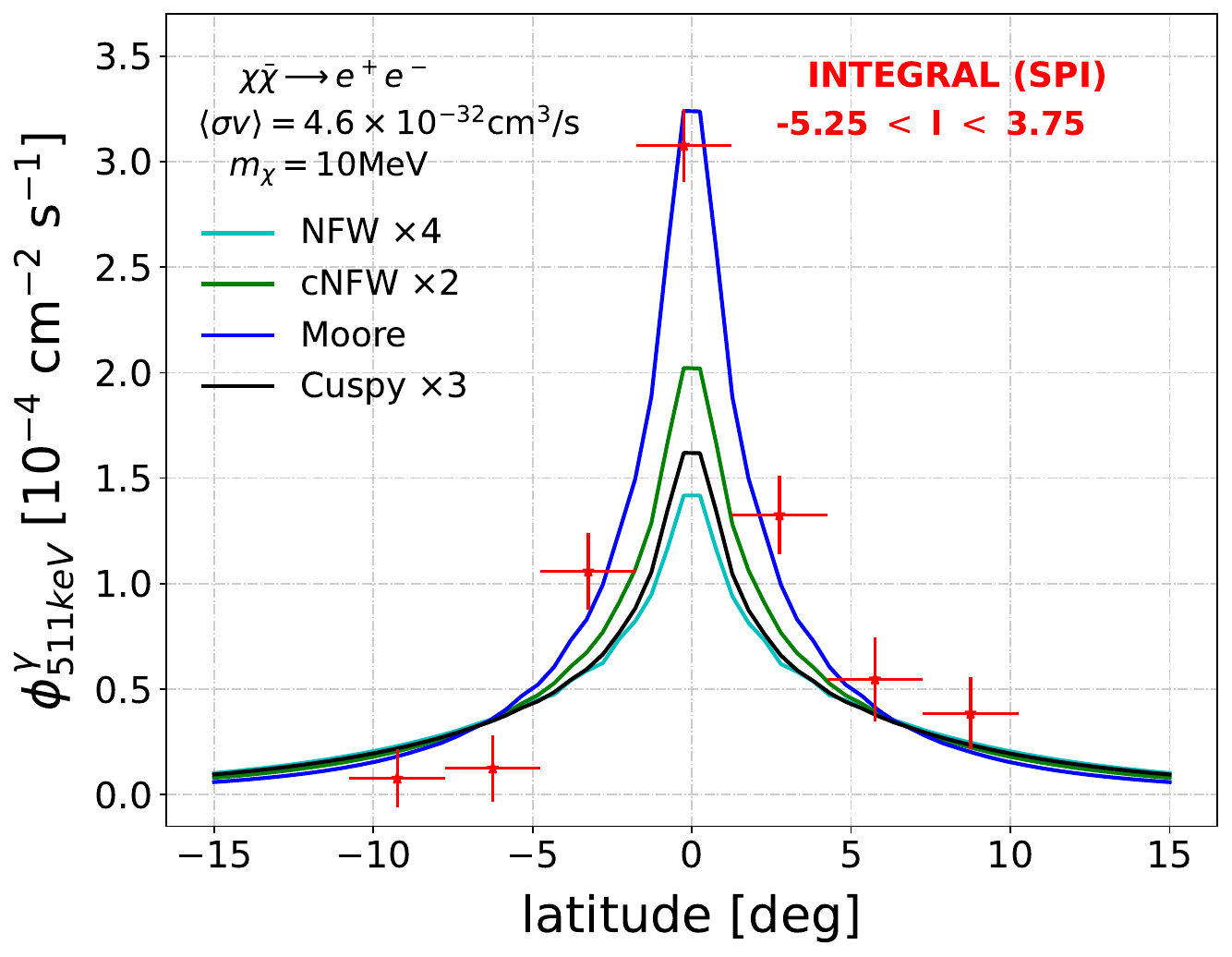}

\includegraphics[width=0.495\linewidth]{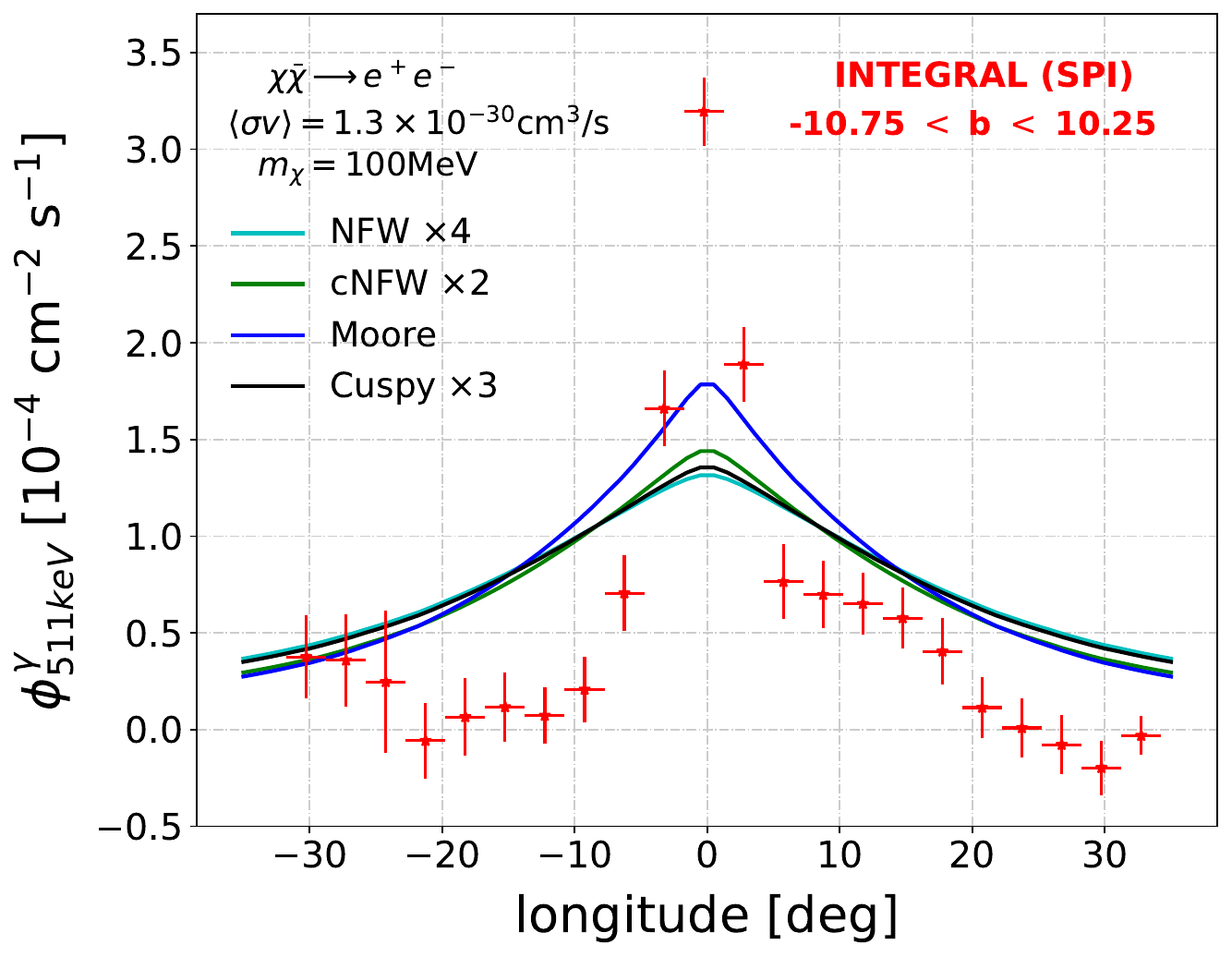}
\includegraphics[width=0.495\linewidth]{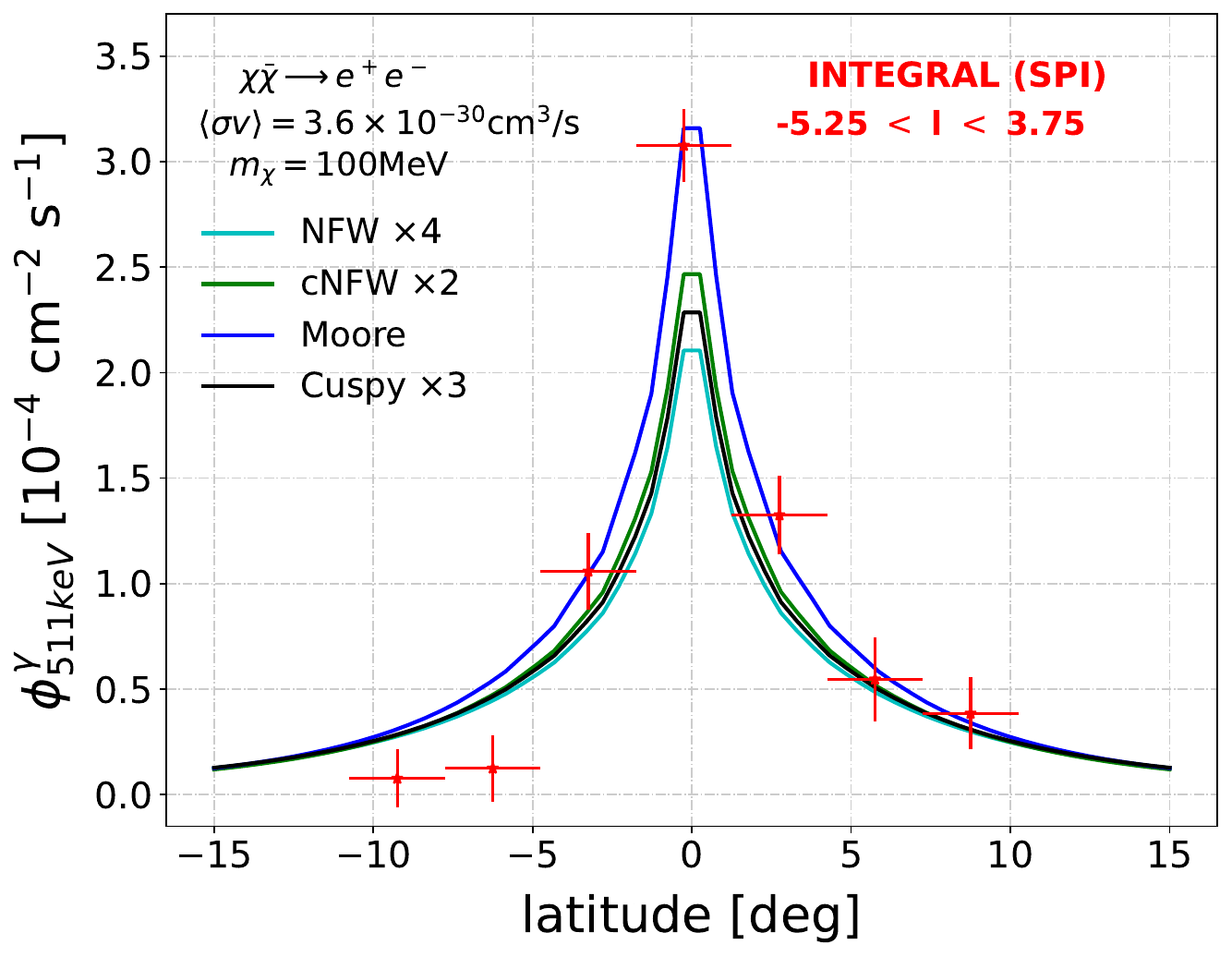}
\caption{Comparison of the predicted diffuse $511$~keV line emission from a low-mass DM particle of mass $m_{\chi}=10$~MeV (top panels) and $m_{\chi}=100$~MeV (bottom panels) annihilating into electron-positron pairs for different DM density profiles.The left panels are the predicted longitude profiles, for latitude bin $-10.75^\circ<b<10.25^\circ$ while the right panels are the latitude profile, for a longitude bin of $-5.25^\circ<l<3.75^\circ$. We fix the annihilation rate to the values shown in the plots and scale the profiles for different masses to facilitate the comparison with the SPI data. We show the NFW profile (cyan), contracted NFW (green), Moore (blue) and cuspy DM profiles.} 
\label{fig:Ann_DMProfs_ee}
\end{figure*}

In this appendix, we report a comparison of the longitude (left panels of Fig.~\ref{fig:Ann_DMProfs_ee}) and latitude (right panels of Fig.~\ref{fig:Ann_DMProfs_ee}) profiles predicted for DM annihilation, assuming different standard Galactic DM distributions, for DM masses of $10$ and $100$~MeV. In particular, we show the predictions obtained for the NFW profile (cyan lines), a contracted NFW profile (c-NFW), like the one that provides the best-fit to the Galactic Center Excess (GCE)~\cite{Ackermann_2017}, which has a slope of $\gamma=1.25$ (green lines), a Moore profile~\cite{Moore_1999} , which has a slope of $\gamma=1.5$ (blue lines), and a cuspy profile, as an example of a distribution that accounts for spikes formed around the central black hole Sgr~A*, taken from Refs.~\cite{Gondolo_1999,Lacroix:2018zmg,Balaji:2023hmy} (black lines).

We defer the full study of the compatibility of the different DM distributions with the SPI observations to a future work. However, this comparison allows us to see how the profiles are sensitive to the different choice of the Galactic DM distribution. In particular, we see that profiles that are not very steep (with slope $\gamma\lesssim 1.25$) are not able to produce a good simultaneous fit to the longitudinal and latitudinal SPI profiles for DM mass exceeding an MeV. However, a steeper profile would be able to provide relatively satisfactory fits for MeV masses. However, we remark that a high slope, like that adopted in the Moore profile, is already in tension with some astrophysical observations. This is because it can predict too much DM in the inner regions of galaxies and galaxy clusters to comply with rotation curve observations \cite{Spekkens:2005ik}, $X$-ray observations of galactic clusters \cite{Walker:2012de}, gravitational lensing observations \cite{Fo_x_2012,Vegetti:2014wza} and N-body simulations of $\Lambda$CDM halos \cite{Navarro:2003ew}. In general, this leads to the conclusion that DM cannot be the dominant source of positrons in the Galaxy, but, instead, it is expected that a combination of different sources is producing the peaked $511$~keV emission profiles that SPI observes.

\section{Annihilation and decay profiles for different channels}
\label{sec:Mass_Prof}

\begin{figure*}[t!]
\includegraphics[width=0.48\linewidth]{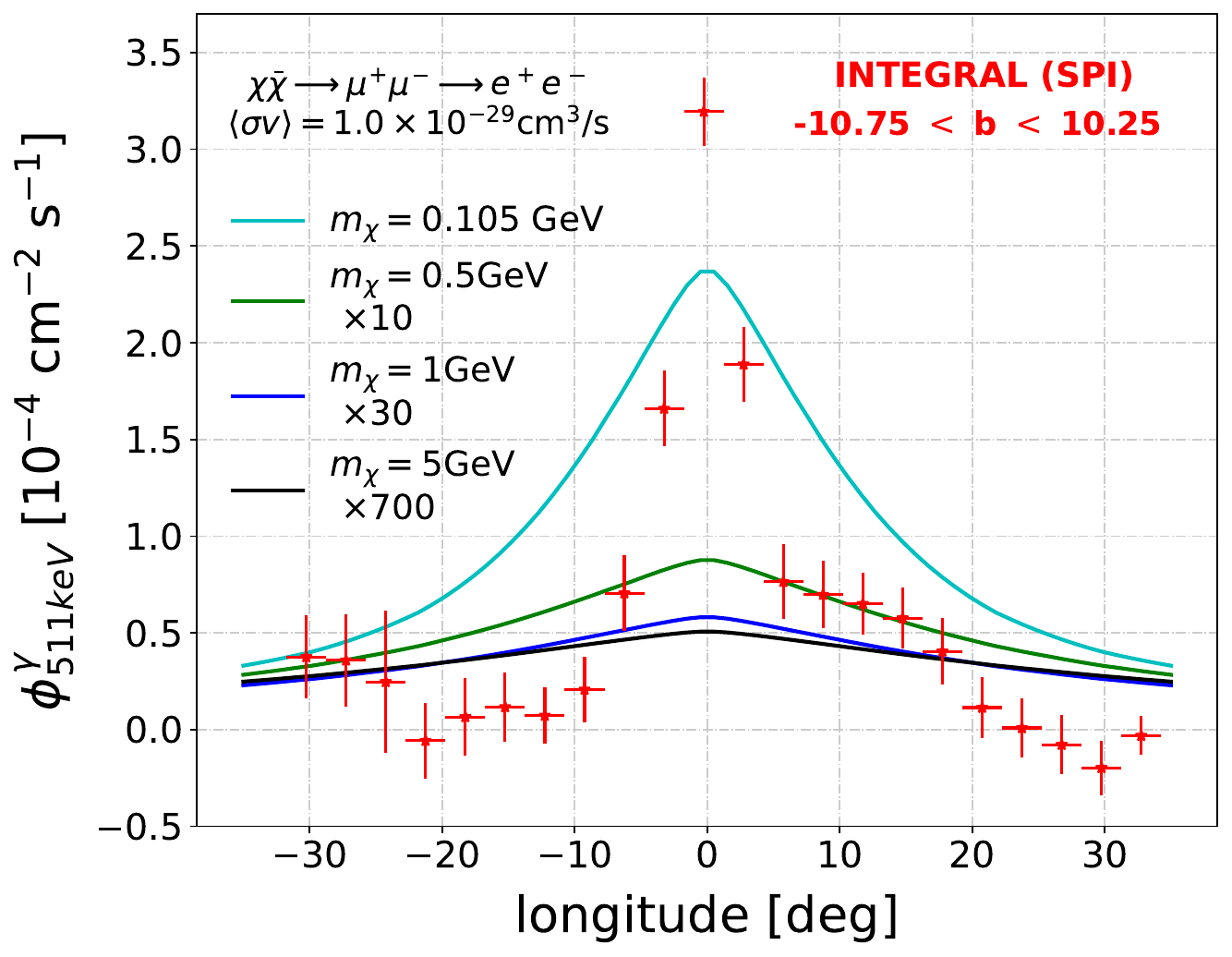}
\includegraphics[width=0.48\linewidth]{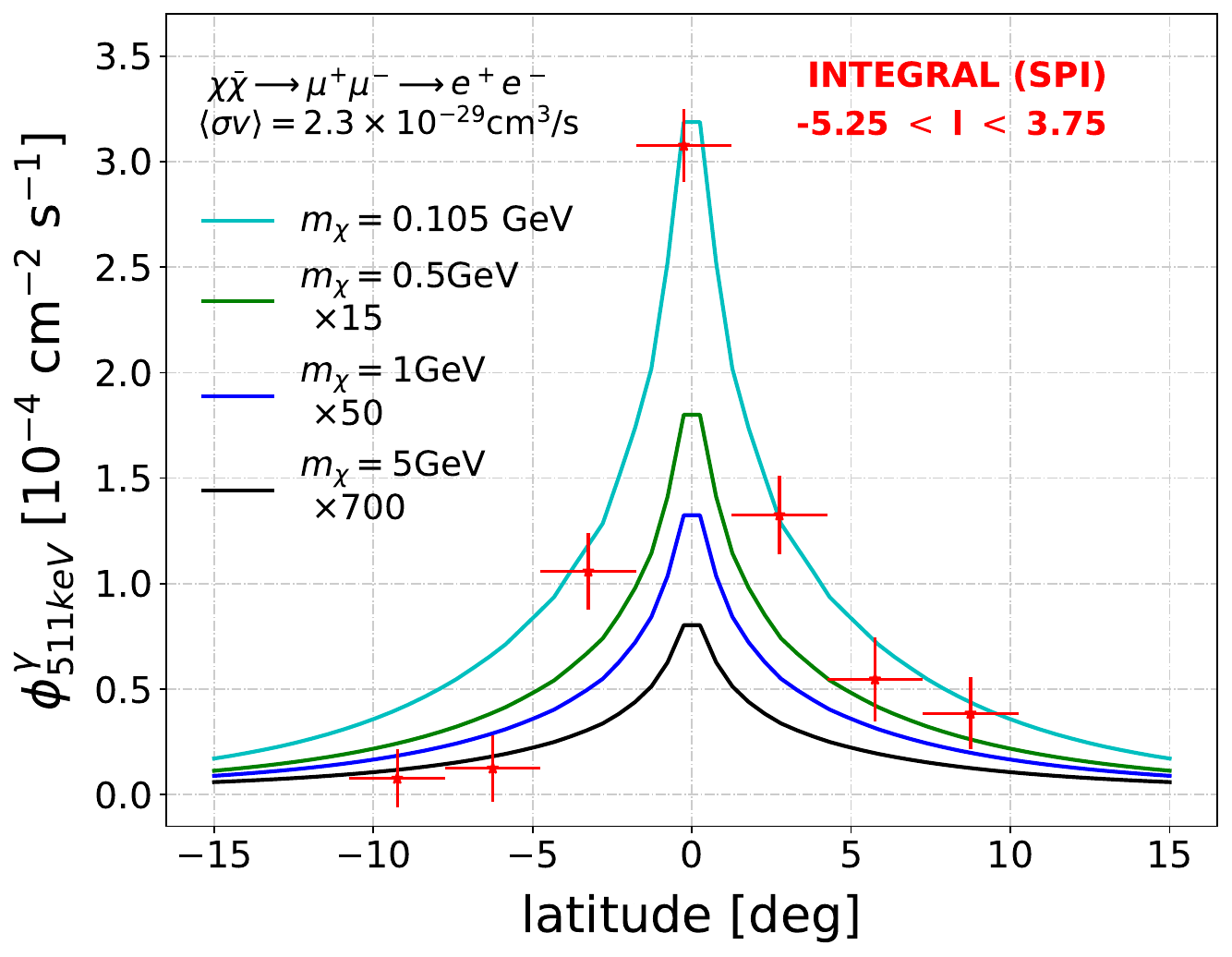}

\includegraphics[width=0.48\linewidth]{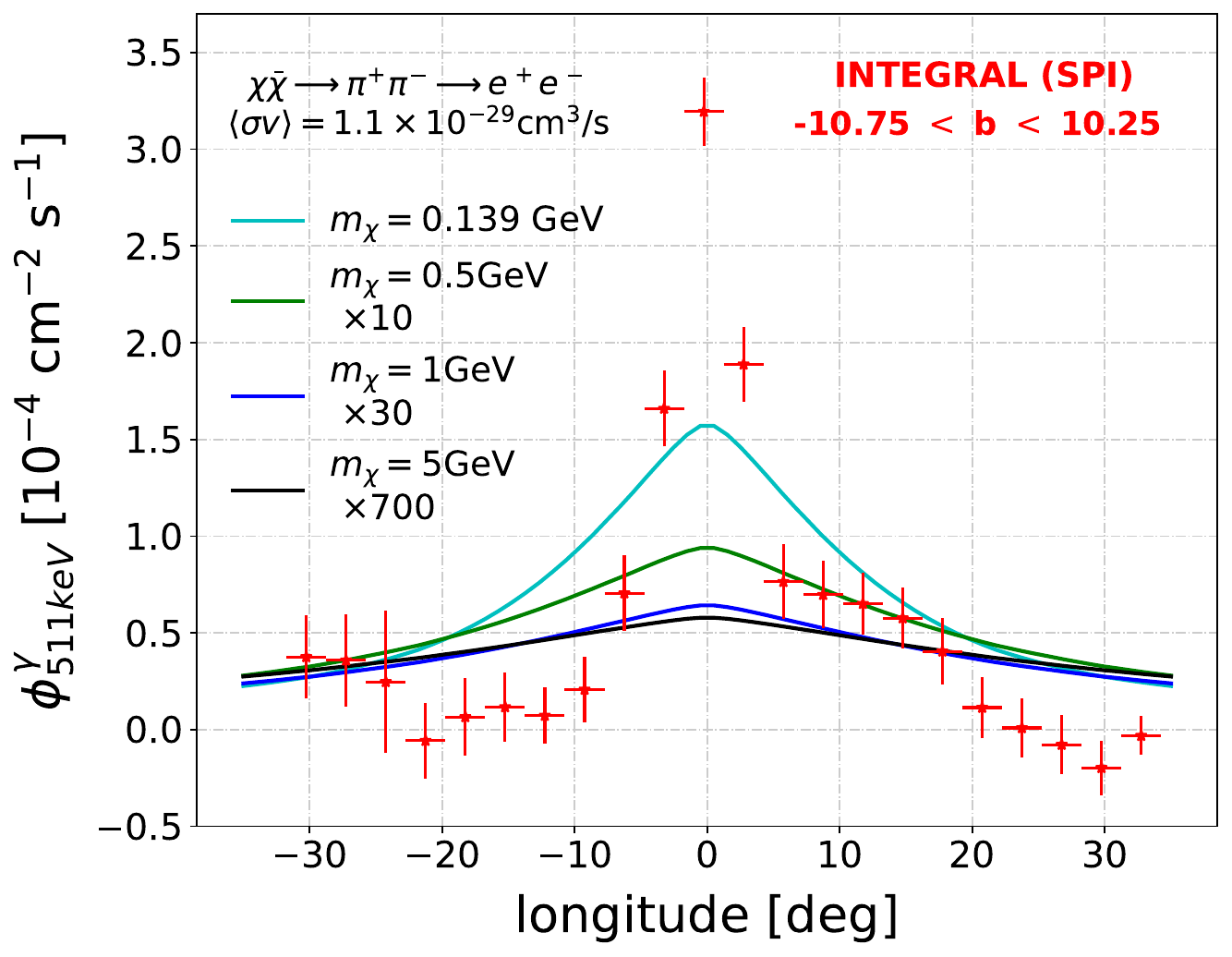}
\includegraphics[width=0.48\linewidth]{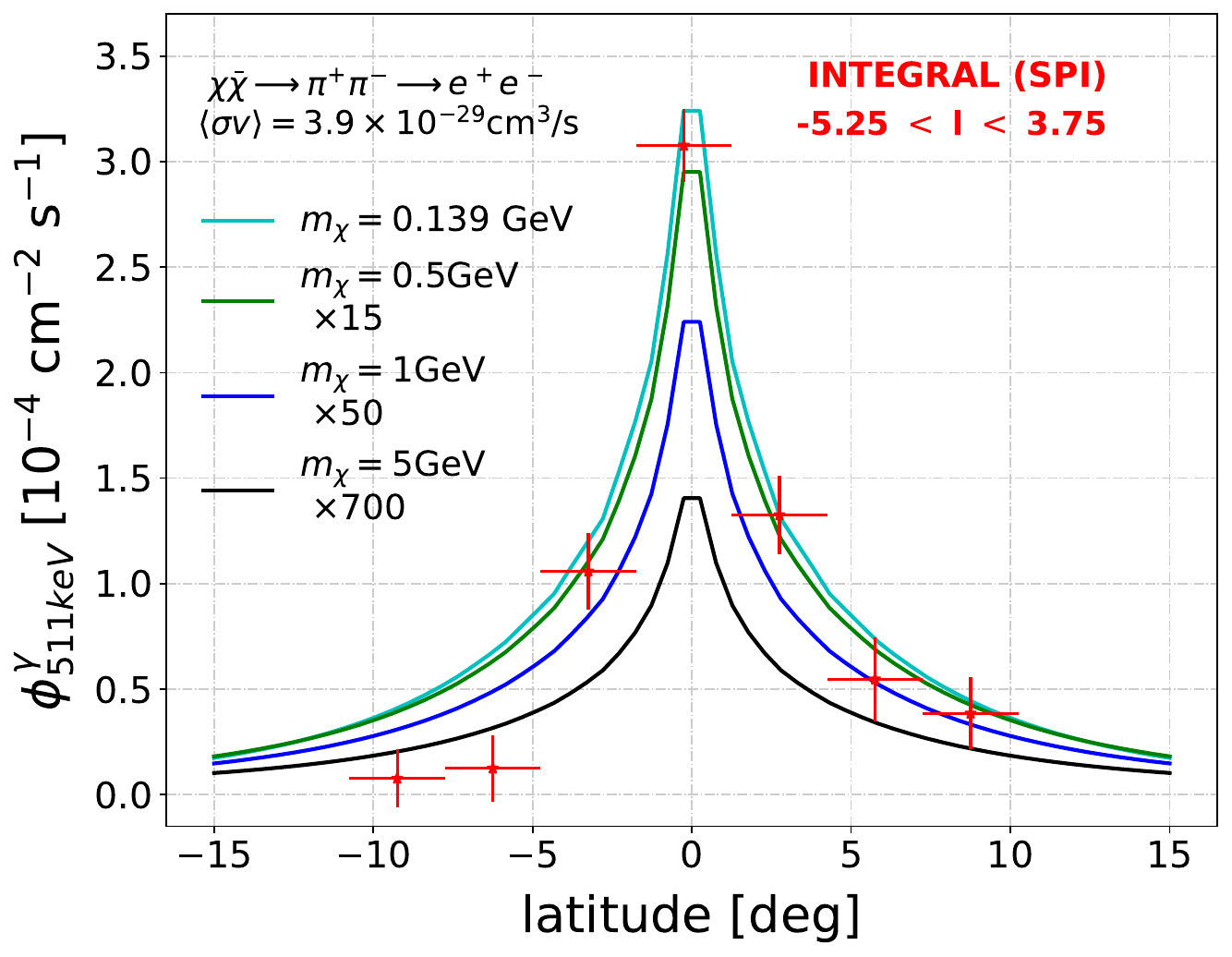}

\includegraphics[width=0.48\linewidth]{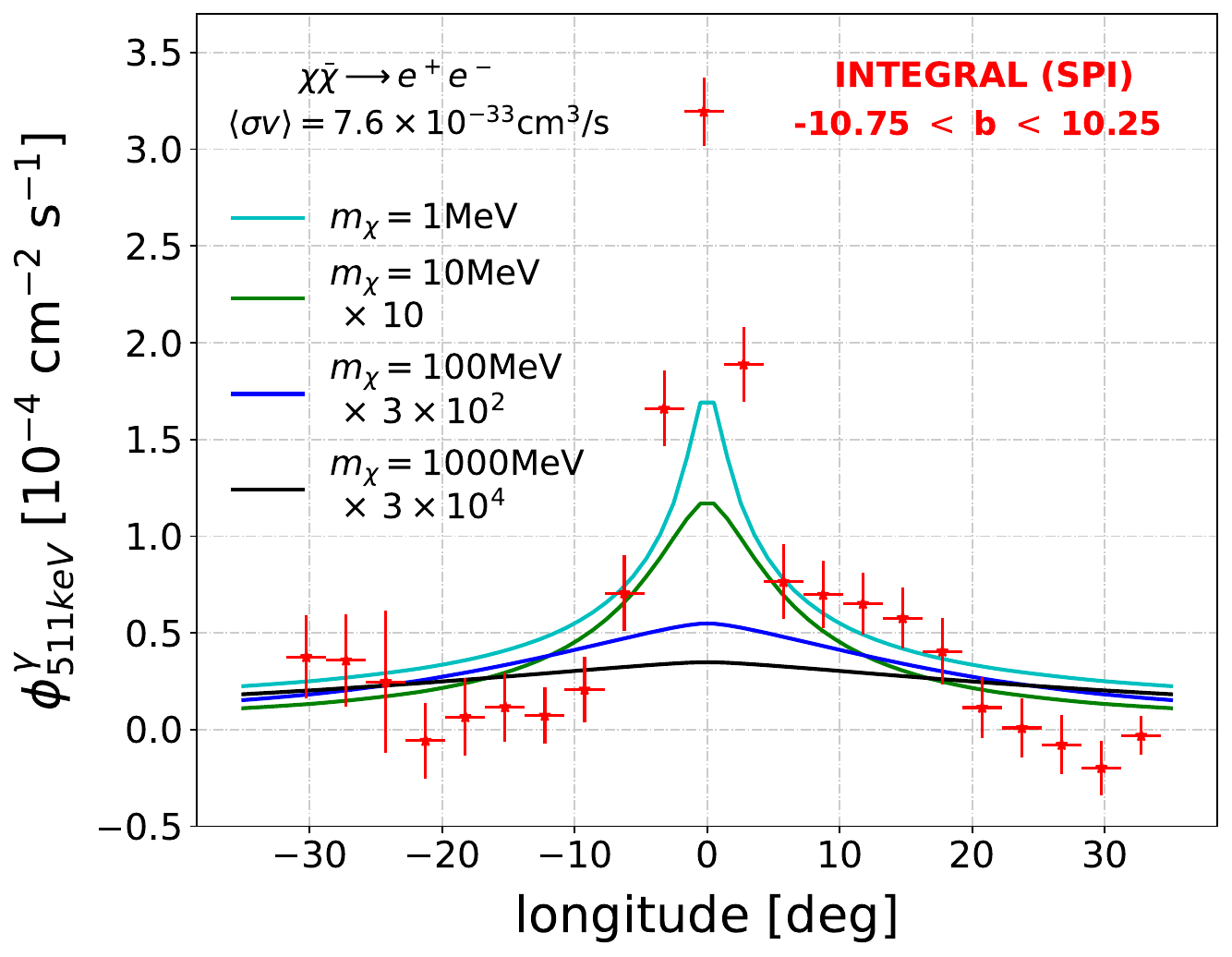}
\includegraphics[width=0.48\linewidth]{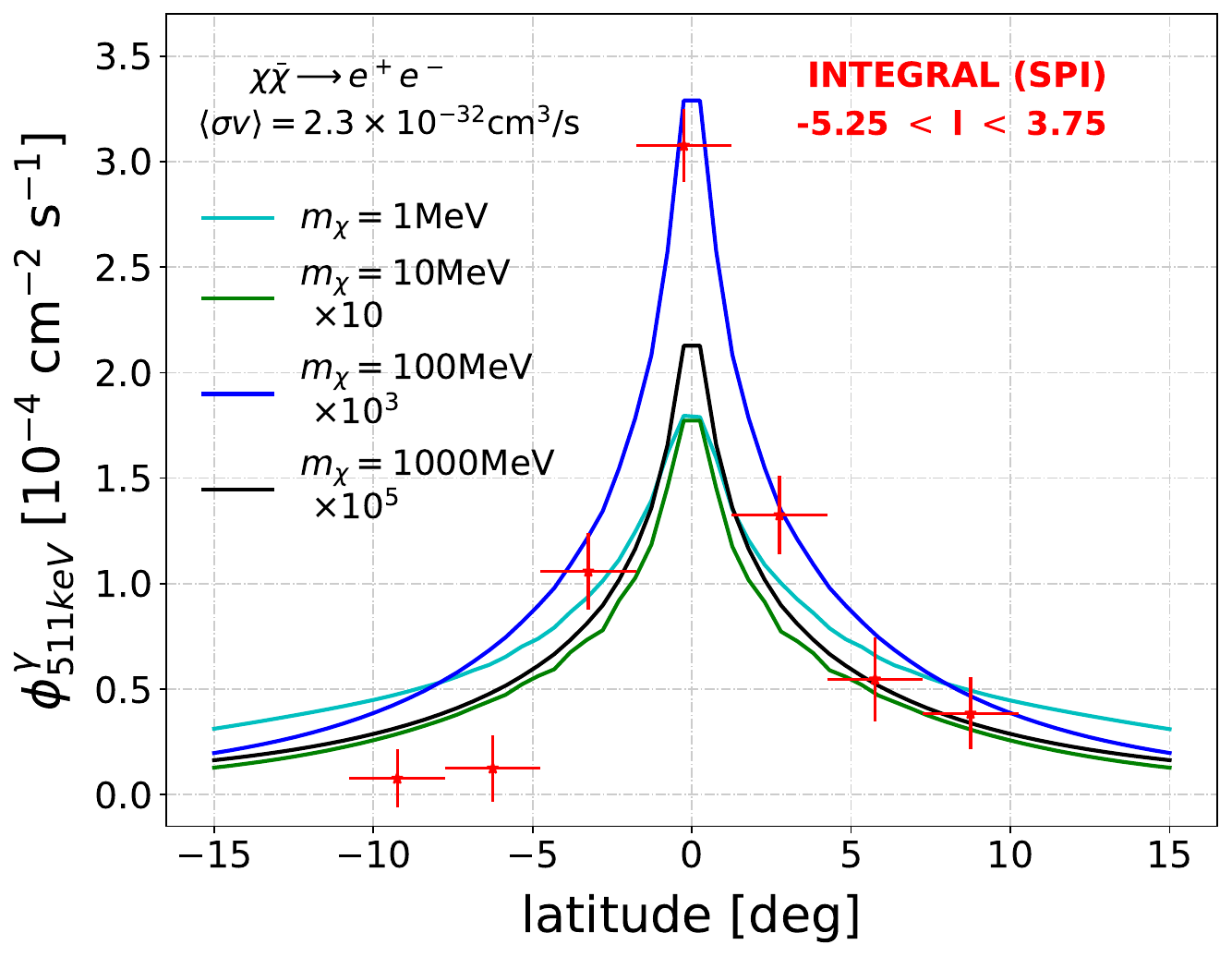}
\caption{Comparison of the predicted longitude (left panels) and latitude (right panels) of the $511$~keV emission from DM annihilation for different DM masses, through the $\mu^+\mu^-$ (top panels), $\pi^+\pi^-$ (middle panels) and direct $e^+e^-$ (bottom panels) channels. The lowest mass for the case of the $\pi^+\pi^-$ annihilation channel is actually $m_{\chi} = 2m_{\pi}\simeq 139$~MeV and the minimum energy shown in the $\mu^+\mu^-$ is $m_{\chi} = 2m_{\mu}\simeq 105$~MeV. We fix the annihilation rate to the values shown in the plots and scale the profiles for different masses to facilitate the comparison with the SPI data. The DM profiles are shown in the same colours and the left and right panels are the same latitude bins as Fig.~\ref{fig:Ann_DMProfs_ee}.}
\label{fig:Ann_Prof_pimu}
\end{figure*}

In this appendix, we report the predicted latitude and longitude profiles of the $511$~keV emission from DM decay and annihilation. In particular, we show in Fig.~\ref{fig:Ann_Prof_pimu} a comparison of profiles obtained assuming DM annihilation through the $\mu^+ \mu^-$ (top panels), $\pi^+\pi^-$ (middle panelbos) and direct $e^+ e^-$ (bottom panels) channels for different DM masses (from twice the mass of the final state, around $100$~MeV, to $5$~GeV, in the case of the $\pi^+ \pi^-$ and $\mu^+ \mu^-$ channels, and from $1$~MeV to $1$~GeV for the direct $e^+e^-$ channel). The left panels illustrate the predicted longitude profiles of the $511$~keV emission and the right panels the predicted latitude profiles. These comparisons allow us to see that the data points corresponding to the higher longitude and latitude angles (in absolute value) are the most constraining ones, which explains why the bounds derived from this dataset are much stronger than those derived from Ref.~\cite{Vincent:2012an} (besides the fact that these measurements are reported after double the duration of event collection).
As we see, the predicted longitude profiles show a clear trend: the lower the mass of the DM particle, the more concentrated the $511$~keV emission around the GC. This is expected, since a lower DM mass means a lower injection energy for the positrons, and the lower the energy of the positrons, the slower they diffuse. Hence lower energy positrons have a distribution more similar to the DM distribution (NFW in these calculations) and lose their energy closer to their injection point. As we pointed out in the previous section, the predictions from a NFW DM distribution do not lead to a $511$~keV emission longitude profile as peaky as that observed by the SPI data even for the low DM masses that are accessible in the direct $e^+e^-$ channel. In turn, as we see from the predicted latitude profiles, masses of around $100$~MeV reproduce the observed SPI latitude profile quite well. We remark that this is due to the scaling that accounts for the decay of the density of free electrons away from the GC, applied for first time in this work. Without accounting for this effect, we observe that the latitude profiles become much more diffuse (i.e. flatter profiles) than what is observed. 
In addition, for the latitude profiles, we see that the same trend wherein lower DM mass leads to a profile more concentrated around the Galactic Plane is observed above some tens of MeV. However, once the energy loss (mainly Coulomb and ionization energy losses) becomes the dominant process during the positron propagation, the $511$~keV emission at the Galactic plane starts to decay quickly, given that these particles suffer from severe energy losses and annihilation in flight losses (since this effect is proportional to the gas density, it does not affect the positrons above and below the Galatic plane). We have investigated that the minimum energy employed in these simulations ($100$~eV) is low enough that our results do not change appreciably for a minimum energy of $1$~eV.

\begin{figure*}[h!]
\centering
\includegraphics[width=0.49\linewidth]{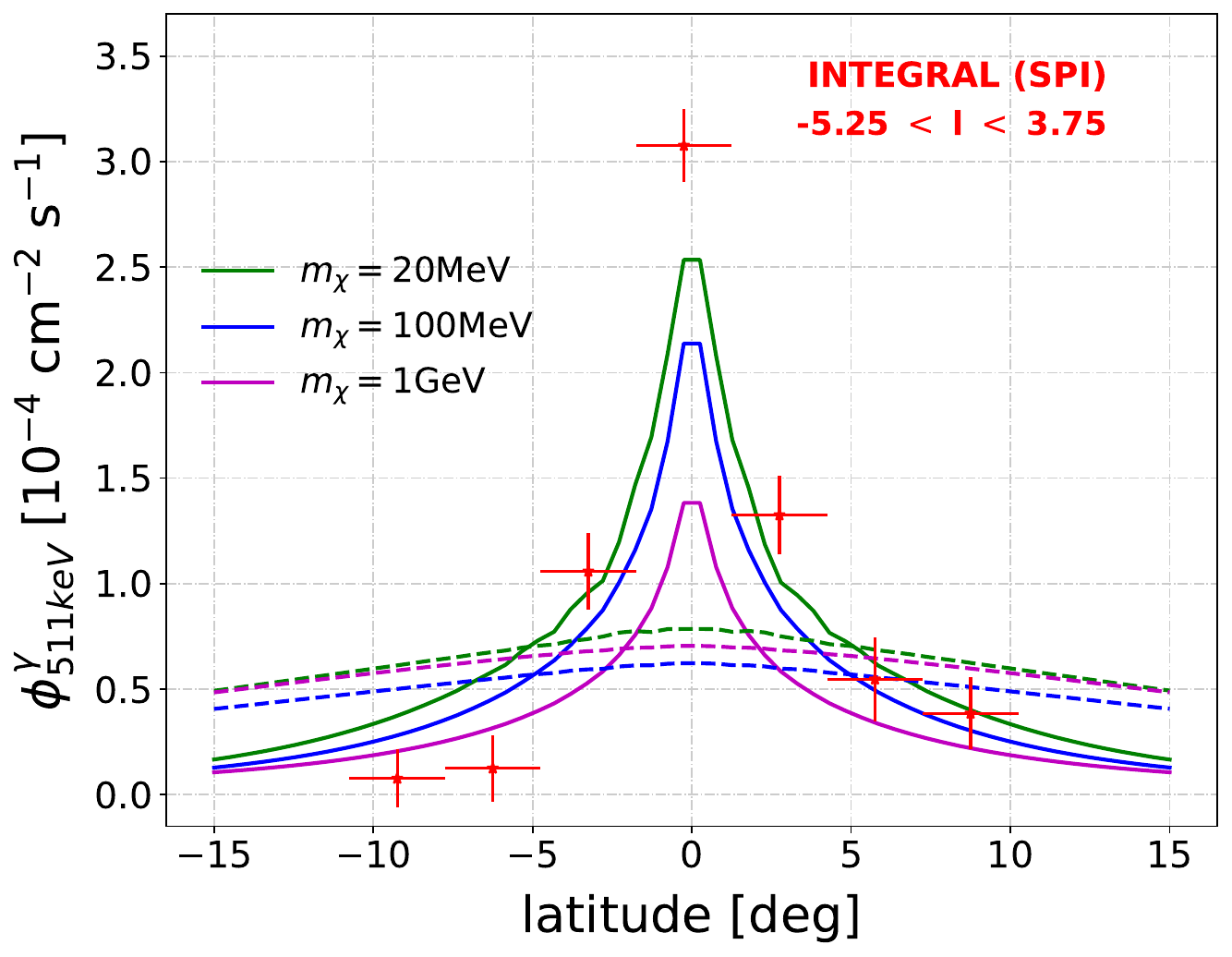}\hspace{0.2cm}
\caption{Comparison of the evaluated latitude profiles accounting for the scaling with the electron density (solid lines) and without considering it (dashed lines).}
\label{fig:Lat_Scaling_Effect}
\end{figure*}

We show a direct comparison of the latitude profile including the density scaling (solid lines) and without including it (dashed lines) for different DM masses in Fig.~\ref{fig:Lat_Scaling_Effect}, normalized to facilitate its visual comparison. This illustrates the importance of accounting for the density scaling to realistically evaluate the latitude profile of the $511$~keV diffuse emission.

\begin{figure*}[h!]
\includegraphics[width=0.48\linewidth]{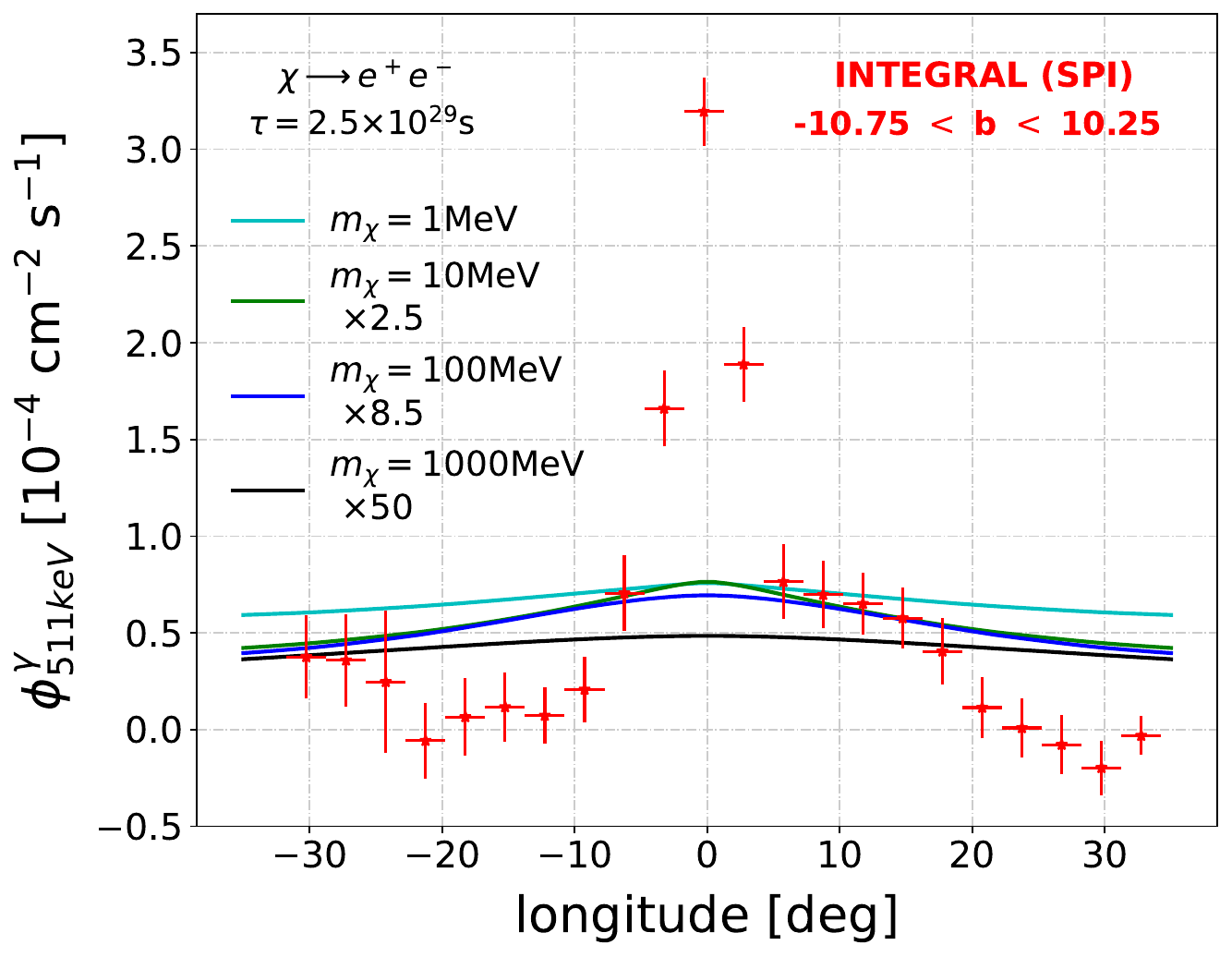}\includegraphics[width=0.48\linewidth]{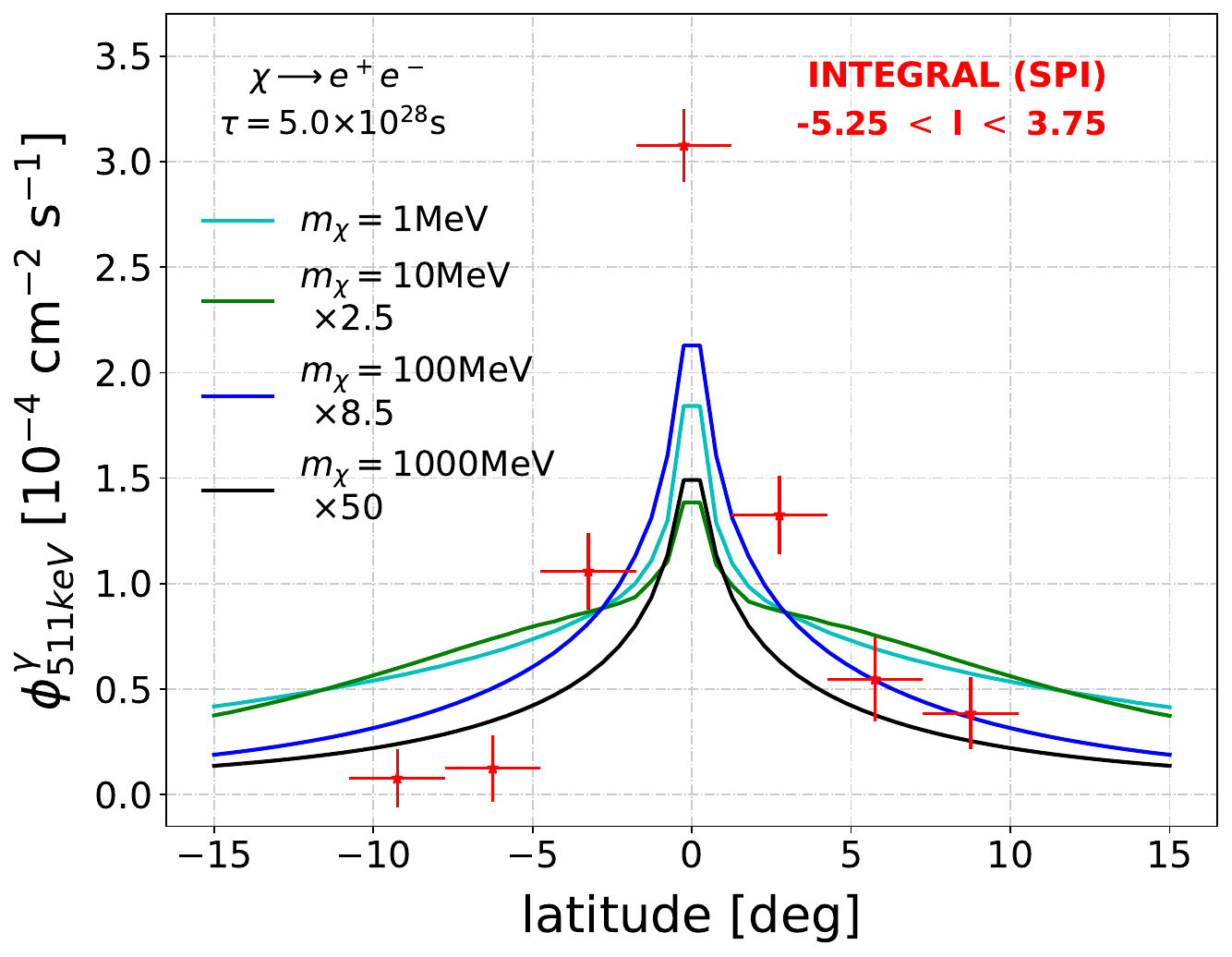} \hspace{0.2cm}
\caption{Similar to Fig.~\ref{fig:Ann_Prof_pimu} but for the case of DM decay into the direct $e^+e^-$ channel. }
\label{fig:Dec_MassComp_ee}
\end{figure*}

Finally, we show in Fig.~\ref{fig:Dec_MassComp_ee} the predicted profiles from decay of DM for masses of $1$~MeV to $1$~GeV, in the direct $e^+ e^-$ channel. As already observed in the past, and discussed above, we observed that both the latitude and longitude profiles of the predicted $511$~keV emission are much flatter (for every mass) than the ones measured by SPI, disfavouring DM decay as the main producer of the emission around the GC. This is simply due to the fact that the total positron emission from decay is proportional to the DM density, while in the case of annihilation the emission is $\propto \rho_{DM}^2$.

\vspace{0.5cm}

\clearpage

\section*{Erratum version}
\subsection*{Abstract}
We correct the results obtained using the diffuse $X$-ray emission from {\sc Xmm-Newton}, in light of new results from Ref.~\cite{Balaji:2025afr}, that demonstrated that the dataset employed was misevaluated. We also update our calculation of the $511$~keV emission from evaporating PBHs, which leads to slightly more conservative constraints as well.
\subsection*{Main text:}
\setcounter{figure}{0}
\renewcommand{\thefigure}{S\arabic{figure}}
\renewcommand{\thesection}{S\Roman{section}}


\title{Erratum: New 511 keV line data provides strongest sub-GeV dark matter constraints}

\author{Pedro De la Torre Luque}\email{pedro.delatorre@uam.es}
 \affiliation{Departamento de F\'{i}sica Te\'{o}rica, M-15, Universidad Aut\'{o}noma de Madrid, E-28049 Madrid, Spain}
\affiliation{Instituto de F\'{i}sica Te\'{o}rica UAM-CSIC, Universidad Aut\'{o}noma de Madrid, C/ Nicol\'{a}s Cabrera, 13-15, 28049 Madrid, Spain}

  \author{Shyam Balaji}
 \email{shyam.balaji@kcl.ac.uk}
 \affiliation{Physics Department, King’s College London, Strand, London, WC2R 2LS, United Kingdom}
\author{Joseph Silk}
\email{silk@iap.fr}
\affiliation{Institut d’Astrophysique de Paris, UMR 7095 CNRS \& Sorbonne Universit\'{e}, 98 bis boulevard Arago, F-75014 Paris, France}
\affiliation{Department of Physics and Astronomy, The Johns Hopkins University, 3400 N. Charles	Street, Baltimore, MD 21218, U.S.A.}
\affiliation{Beecroft Institute for Particle Astrophysics and Cosmology, University of Oxford, Keble	Road, Oxford OX1 3RH, U.K.}



\section*{}
\vspace{-0.7cm}
In the original paper~\cite{DelaTorreLuque:2023cef}, our calculation of the 511 keV flux inadvertently overcounted the number of positrons contributing to positronium formation. By integrating the positron flux over \textit{all} energies, we included high-energy positrons that typically escape or annihilate in-flight before thermalizing, and therefore do not produce 511 keV photons in the region studied. As a result, 511 keV flux was over predicted, and the required cross sections/decay lifetimes were quoted too stringently.
More precisely, our expression for the total diffuse positron flux ($\phi_e$ in Eq.~(3) of \cite{DelaTorreLuque:2023cef}) was defined as the integral of the positron flux over all positron energies. Instead, as shown in Ref.\cite{laTorreLuquePedro:2024est}, a correct evaluation of the 511 keV flux requires taking the flux of positrons at the \textit{thermal energy} of the interstellar gas where positronium forms, which, assuming warm gas, can be taken to be $E_{th}\approx100$ eV (see Ref.~\cite{laTorreLuquePedro:2024est} for further discussion). The corrected calculation of the $511$ keV flux per solid angle is the following
\begin{equation}
\frac{d\phi_{\gamma}^{511}}{d\Omega}=2k_{ps}\int ds\,s^{2}\frac{\epsilon_e(x_{s,b,l},y_{s,b,l},z_{s,b,l})}{4\pi s^{2}},
\end{equation}
where $k_{ps}=1/4$ is the fraction of positronium decays corresponding to (singlet) para-positronium states) contributing to the  $511$~keV line signal, the factor of $2$ accounts for the production of two photons in these decays, and $\epsilon_{\rm e}$ is the energy-integrated emissivity of 511 keV photons, defined as:
\begin{equation}
\epsilon_{\rm e} = \frac{d\phi_{e}}{d\Omega} (E_\textrm{th}) \cdot n_e \cdot \sigma^\textrm{ps}(E_\textrm{th})\,,
\end{equation}
where $\frac{d\phi_{\rm e}}{dE}(E_\textrm{th})$ is the positron flux at the thermal energy, the positronium formation cross sections $\sigma^\textrm{ps}$ are the charge-exchange cross section with hydrogen~\cite{JeanP_2009}. The electron density, $n_e$, is set to $1$~cm$^{-3}$ in the Galactic plane and having the vertical dependence of free electron density as described in the original paper.

This change in our estimations makes the spatial profile of the 511 keV line follow almost exactly our previous evaluation, but it significantly varies the normalization (i.e. the annihilation or decay rates) of the predicted signals. 
In fact, we generally find larger deviations in the normalization of the signals for larger DM masses -- a $1$~MeV DM particle requires a factor $\sim10$ larger cross section than in our previous estimation (in case of decay, a factor $35$ lower decay rate), at $m_{\chi}=10$~MeV a factor $\sim120$ ($1/100$ for decay), at $m_{\chi}=100$~MeV a factor $\sim400$ ($1/450$ for decay), and at $m_{\chi}=1000$~MeV a factor $\sim10^3$ ($1/4000$ for decay). These factors are roughly the same for the different annihilation/decay channels. For brevity, we do not repeat every numerical value quoted in the original text, the updated constraints should be read with the revised normalization factors given above and in Figs. 1–2.

\begin{figure}[t!]
\centering
\includegraphics[width=0.99\linewidth]{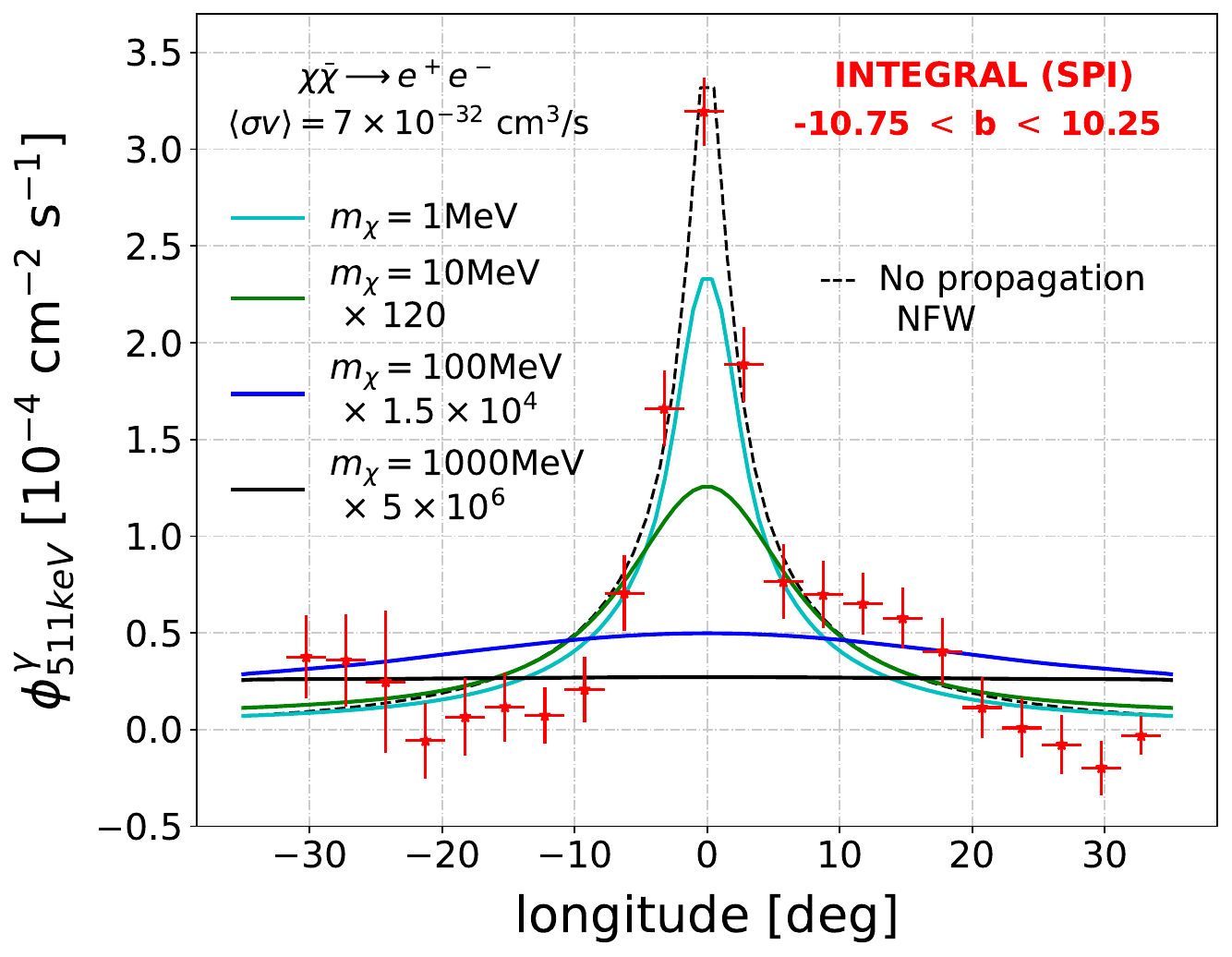}
\caption{Diffuse $511$~keV emission longitude profile in the latitude bin $-10.75^\circ<b<10.25^\circ$ from direct DM annihilation into electron-positron pairs. We show DM masses ($m_\chi$) ranging from $1$ (cyan), $10$ (green), $100$ (blue) and 1000~MeV (black) respectively, and compare to the expected signal with no propagation (dashed). To facilitate the comparison with SPI data (shown in red), the emission for each mass is scaled by the factor indicated in the legend. We consider a typical NFW DM distribution.} 
\label{fig:Ann_MassComp_ee_Err}
\end{figure}

\begin{figure*}[t!]
\includegraphics[width=0.49\linewidth]{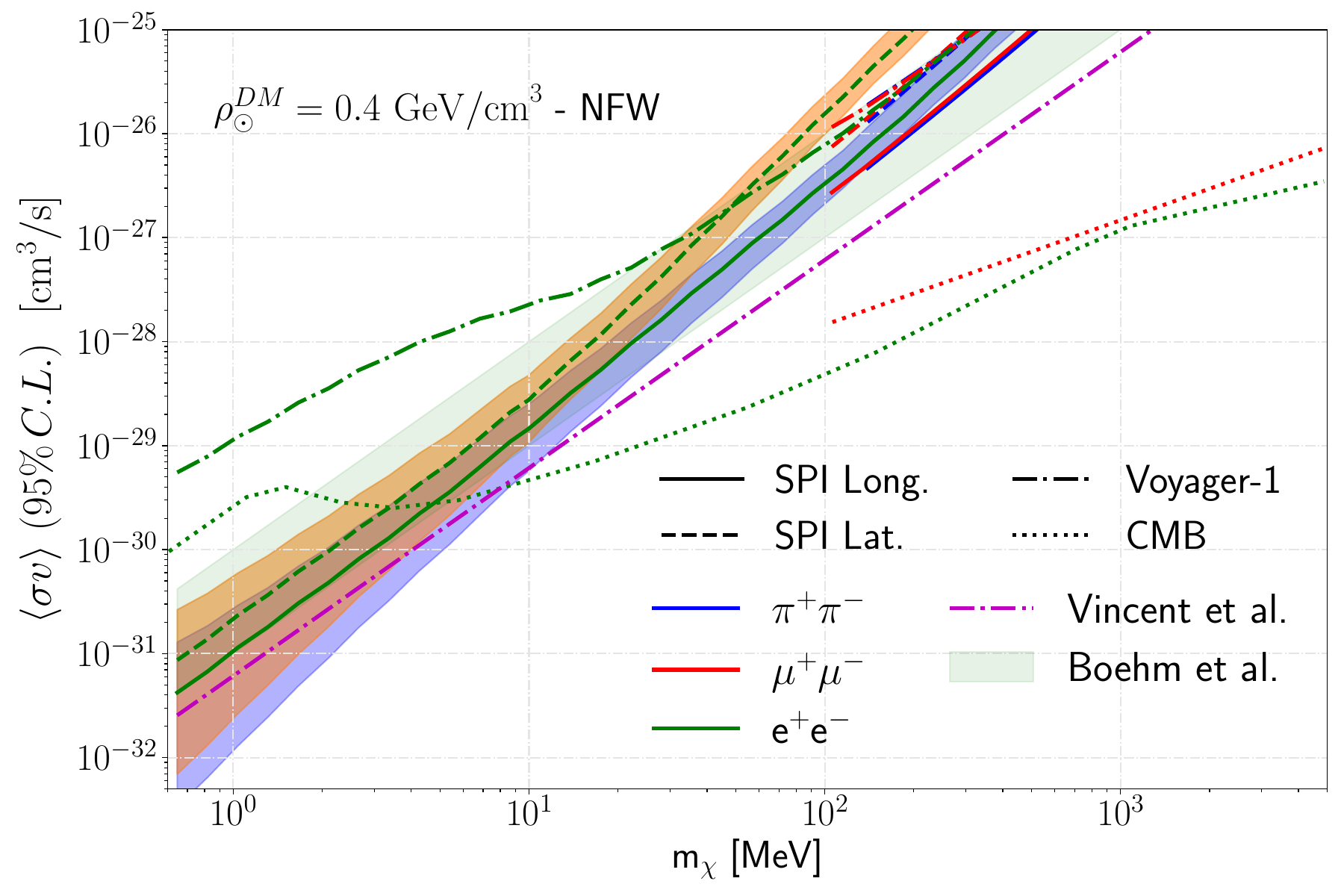}
\includegraphics[width=0.49\linewidth]{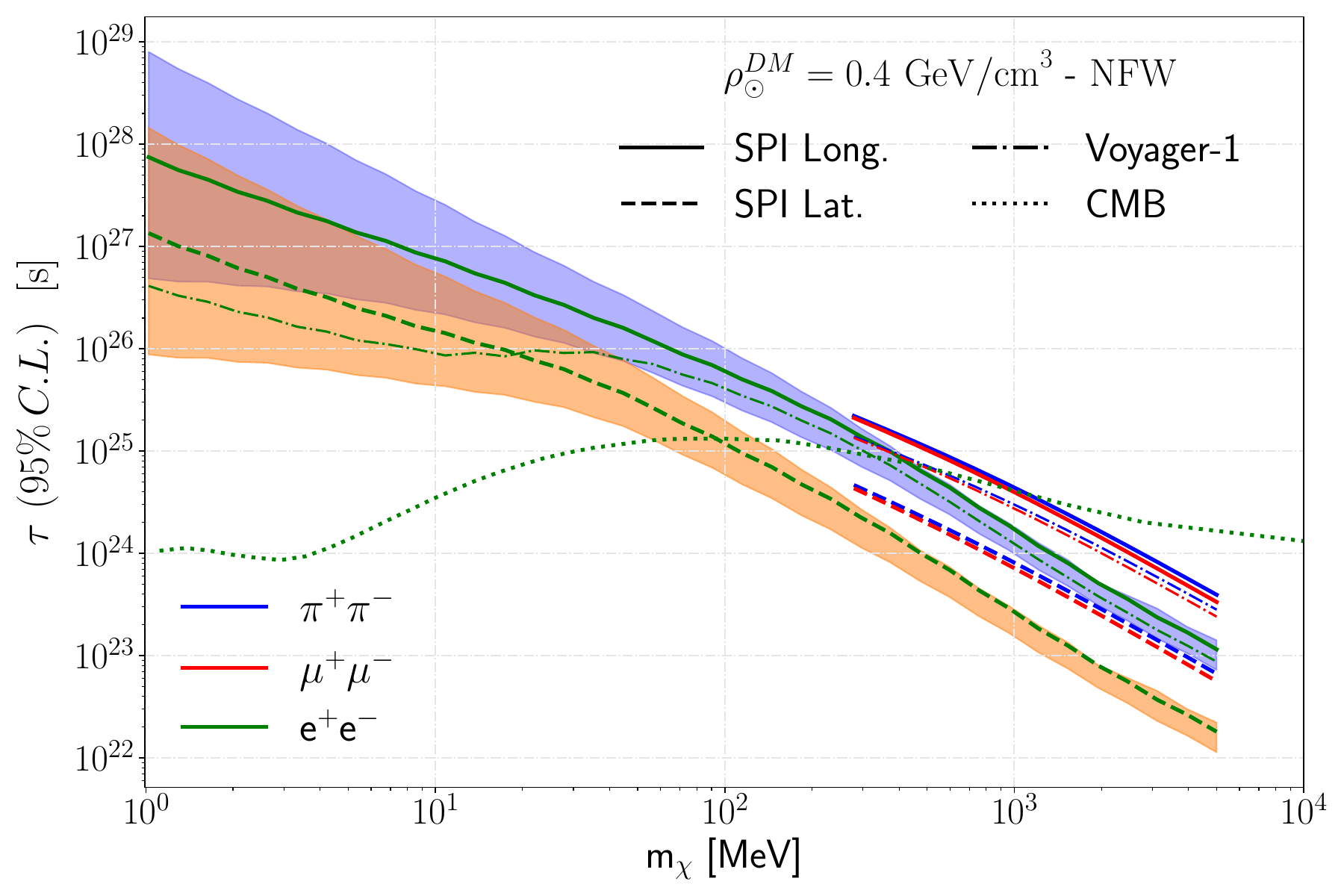}
\caption{Comparison of the $95\%$ confidence bounds on annihilating (left panel) or decaying (right panel) DM derived in this work (thick dashed for {\sc SPI Lat} and solid lines for {\sc SPI Long}, respectively) with other existing constraints. We show the CMB bounds for annihilating DM from Slayter~\cite{Slatyer:2015jla} and Lopez-Honorez et al.~\cite{Lopez-Honorez:2013cua} and for decaying DM from Ref.~\cite{Liu_2016} (dotted lines), the previous 511 keV line fits from Vincent et. al~\cite{Vincent:2012an} assuming a contracted NFW profile with $\gamma=1.2$ (red dashed line), and from {\sc Voyager 1}~\cite{DelaTorreLuque:2023olp} (dot-dashed lines). In both panels, we show the $e^+e^-$ (green), $\mu^+\mu^-$ (red) and $\pi^+\pi^-$ (blue) channels, respectively. Uncertainties from the propagation setup employed are shown as an orange and a blue bands, for the latitude and longitude profiles, respectively.}
\label{fig:Limits_Err}
\end{figure*}

We show the updated version of Fig.~1 in the original paper~\cite{DelaTorreLuque:2023cef} in Fig.~\ref{fig:Ann_MassComp_ee_Err}. As it can be seen, the main difference with respect to the original version is in the annihilation rate required to match the observed fluxes.

We provide in Fig.~\ref{fig:Limits_Err} the updated DM bounds for annihilation (left panel) and decay (right panel) with this updated implementation. Note that these panels do not contain the XMM-Newton bounds, previously shown. This is because an updated analysis of the XMM-Newton diffuse data~\cite{Balaji:2025afr} resulted in constraints that are too weak, in comparison to the ones shown in Fig.~\ref{fig:Limits_Err}.

The signal morphology is essentially unchanged, and the 511 keV line continues to provide highly competitive constraints on sub-GeV dark matter. The corrected implementation affects only the normalization of the limits: cross sections required for annihilation are larger and lifetimes are smaller than previously quoted. This systematically weakens the numerical bounds, but they remain among the strongest astrophysical constraints on light dark matter, particularly at low masses.


\bibliographystyle{aasjournal}
\bibliography{references.bib}

\end{document}